\documentclass[11pt,a4paper]{article}

\usepackage{jheppub}
\usepackage{subcaption} % for \subfigure
\usepackage{bm} % for \bm
\usepackage[section]{placeins} % for \FloatBarrier
\usepackage{multirow,makecell}

\newcommand{\comment}[1]{}

\title{From strange-quark tagging to fragmentation tagging with machine learning}

\begin{document}

\author{Yevgeny Kats}
\author{and Edo Ofir}
\affiliation{Department of Physics, Ben-Gurion University, Beer-Sheva 8410501, Israel}
\emailAdd{katsye@bgu.ac.il}
\emailAdd{edoof@post.bgu.ac.il}

\abstract{We apply advanced machine learning techniques to two challenging jet classification problems at the LHC. The first is strange-quark tagging, in particular distinguishing strange-quark jets from down-quark jets. The second, which we term fragmentation tagging, involves identifying the fragmentation channel of a quark. We exemplify the latter by training neural networks to differentiate between bottom jets containing a bottom baryon and those containing a bottom meson. The common challenge in these two problems is that neither quark lifetimes and masses nor parton showering provide discriminating tools, making it necessary to rely on differences in the distributions of the hadron types contained in each type of jet and their kinematics. For these classification tasks, we employ variations of Graph Attention Networks and the Particle Transformer, which receive jet and all constituent properties as inputs. We compare their performance to a simple Multilayer Perceptron that uses simple variables. We find that the more sophisticated architectures do not improve $s$-quark versus $d$-quark jet differentiation by a significant amount, but they do lead to a significant gain in $b$-baryon versus $b$-meson jet differentiation.}

\maketitle

\section{Introduction}
\label{sec: Introduction}

Jets, which are collimated sprays of particles, are among the most ubiquitous objects produced at the Large Hadron Collider (LHC) at CERN and other high-energy colliders. Determining a jet's origin is often crucial for deciphering the underlying physical process that occurred in the collision. In other cases, one would like to focus on particular hadrons that might have been produced, with their decay products contained in the jet. However, these tasks are complicated by the stochastic nature of the processes governing the evolution of the original particle into a jet, which typically ends up containing tens of particles. Moreover, due to the limitations of detectors, only certain properties of the jet constituents can be measured. Machine learning (ML) tools are therefore a natural fit for analyzing jet data. They have been already proven effective in many such tasks (e.g., refs.~\cite{Mondal:2024nsa,Kasieczka:2019dbj}).

We are interested in strange-quark tagging as part of a broader effort to distinguish between different types of quarks, gluons, and other objects that produce jets. In many cases, various established tagging techniques exist. Bottom and charm quarks travel several millimeters within the detector because of their long lifetimes. Their decays produce secondary vertices, aiding in their recognition and differentiation from other particles~\cite{ATLAS:2022qxm, sirunyan2018identification}. Another example involves distinguishing gluon from light-quark ($u,d,s$) jets. This is done by noting that gluon jets are usually wider, have a larger number of constituents, and exhibit more uniform energy fragmentation~\cite{kogler2019jet}. Similarly, up and down-quark jets can be partially distinguished using the $p_T$-weighted track charge~\cite{krohn2013jet}. Moreover, in scenarios where a heavy hadronically decaying particle (e.g., a $W$, $Z$ or Higgs boson or a top quark) is boosted, such that its decay products are collimated into a single jet, it can be distinguished from quark/gluon jets by using the jet mass or substructure variables (e.g., refs.~\cite{thaler2011identifying,plehn2012top}). Distinguishing between strange and down-quark jets, on the other hand, remains a challenge since these quarks have small masses and identical QCD and electromagnetic interactions, so the only difference between them is found in their hadronization and subsequent decay processes.

Simple methods for strange-quark tagging were implemented by DELPHI~\cite{DELPHI:1994aml,DELPHI:1999mkl} and OPAL~\cite{OPAL:1997tsq} at LEP and by SLD~\cite{SLD:2000jop} at SLAC in measurements of the strange-quark forward-backward asymmetry in $e^+e^-$ collisions near the $Z$ pole. These methods relied on the fact that the particle with the highest energy in a jet tends to carry the flavor of the primary quark. Therefore, strange jets can be characterized by them containing an energetic charged or neutral kaon or a $\Lambda$ baryon. More recently, a combination of such inputs was proposed for identifying Higgs boson decays to $s\bar s$ in a future $e^+e^-$ collider~\cite{duarte1811probing}. A recurrent neural network (RNN) was proposed for the same task in ref.~\cite{Albert:2022mpk}. And a short while ago, a transformer-based neural network was proposed for jet flavor tagging, including strange-quark tagging, for the FCC-ee~\cite{Blekman:2024wyf}.

Our focus will be on the multipurpose LHC detectors ATLAS~\cite{ATLAS:2008xda} and CMS~\cite{CMS:2008xjf}, where the strange-tagging challenge is exacerbated by the inability to determine the identity of charged hadrons. This implies, in particular, that charged kaons ($K^\pm$) cannot be distinguished from charged pions and protons. Additionally, for jet transverse momentum ($p_T$) above a few tens of GeV, most of the energetic short-lived neutral kaons ($K_S$) and $\Lambda$ baryons reach the hadronic calorimeter (HCAL) without decaying, like the long-lived neutral kaons ($K_L$), which makes them indistinguishable from neutrons or a collection of soft neutral hadrons. Still, the presence of these energetic neutral strange hadrons in $s$-quark jets leads to an increased energy fraction, on average, deposited in the HCAL, while $d$-quark jets have a greater energy fraction deposited in the electromagnetic calorimeter (ECAL) due to energetic neutral pions decaying to photons. However, the distributions of these quantities for the two classes of jets overlap significantly, which limits their discriminating power.

Several studies on possible $s$-tagging strategies for ATLAS and CMS have been reported in the literature~\cite{erdmann2020tagger,nakai2020strange,erdmann2021maximum,zeissner2021development}. Ref.~\cite{erdmann2020tagger} considered processing the tracks in the jet with a Long Short-Term Memory (LSTM) RNN. Refs.~\cite{erdmann2021maximum,zeissner2021development} extended this study to use calorimeter information, again with LSTM RNNs. Ref.~\cite{zeissner2021development} also implemented a feedforward neural network (FNN) with multiple properties of the jet and of reconstructed $K_S$ and $\Lambda$ hadrons as inputs. The resulting $s$ tagger was calibrated on ATLAS $t\bar t$ samples with hadronic $W$ decays, and then used to constrain the CKM matrix elements $|V_{ts}|$ and $|V_{td}|$ by analyzing data from top-quark decays. Ref.~\cite{nakai2020strange} considered Boosted Decision Tree (BDT) classifiers with whole-jet energy fractions as inputs, as well as Convolutional Neural Networks (CNNs) applied to jet images, and found the CNNs to outperform the BDTs by a small amount. In all these cases, the discriminating power was found to be quite limited, with AUC scores not exceeding $0.64$.

In this paper, we make another attempt to tackle the difficult problem of strange-jet tagging by employing different and more sophisticated neural network (NN) architectures that utilize \emph{attention mechanisms}. One is a variant of the Graph Attention Network (GAT)~\cite{Velickovic:2017lzs,Brody:2021dbs}, which is a powerful Graph Neural Network (GNN)~\cite{shlomi2020graph,DeZoort:2023vrm} architecture. Another NN we use is based on the attention mechanisms utilized in the famous \emph{transformer} architecture~\cite{vaswani2017attention,DBLP:journals/corr/abs-2103-17239}. Transformers were shown to outperform GNNs and CNNs in sequence transduction problems~\cite{vaswani2017attention} and they are at the heart of state-of-the-art artificial intelligence applications, such as ChatGPT~\cite{openai2024chatgpt}. In collider physics, the recently introduced Particle Transformer (ParT)~\cite{qu2022particle} (see also refs.~\cite{ATL-PHYS-PUB-2023-032,Usman:2024hxz,Wu:2024thh,CMS-PAS-HIG-23-012} for more sophisticated versions) was shown to be very effective in many jet classification tasks, and we will closely follow its architecture. Similar architectures are also being explored by the ATLAS and CMS collaborations for various jet tagging tasks~\cite{Duperrin:2023elp,ATL-PHYS-PUB-2023-021,ATL-PHYS-PUB-2023-032,CMS-DP-2022-050,CMS-PAS-HIG-23-012}.

The second problem that we will address with the same tools is \emph{fragmentation tagging}. By this we mean determining the fragmentation channel of a quark, e.g., identifying the type of $b$ hadron that was present in a $b$ jet. The motivation for developing fragmentation taggers comes from several directions. One is measurements of parton fragmentation functions (FFs), which describe how partons transform into hadrons~\cite{metz2016parton,albino2008akk}. The FFs, which are determined by nonperturbative QCD matrix elements, describe the probability of a parton producing a specific hadron carrying a certain fraction of the parton momentum. Measuring FFs is useful for improving our understanding, or at least the description, of QCD dynamics, as well as for tuning Monte Carlo generators. While FF measurements can be done using specific clean decay channels of the corresponding hadrons (see, e.g., refs.~\cite{seuster2006charm,CMS-PAS-TOP-18-012}), it is interesting to ask whether ML methods can help doing more inclusive measurements. Such measurements might be particularly useful at high $p_T$, where statistics is limited.

Another motivation for fragmentation tagging is analyses targeting jets with (decay products of) specific hadrons, where jets with other hadrons form a background. One such case is the proposed measurements of $b$-quark polarization and/or $b\bar b$ spin correlations in various samples in ATLAS and CMS~\cite{galanti2015heavy, kats2023prospects, Afik:2024uif}. These proposals rely on reconstructing semileptonic decays of $\Lambda_b$ baryons in $b$ jets. A fragmentation tagger could help reduce the large background from semileptonic decays of $B$ mesons.

We will demonstrate fragmentation tagging using the example of distinguishing between $b$ jets containing a $b$ baryon vs.\ those containing a $b$ meson. Various other definitions of classes (e.g., focusing on specific hadrons as signal, distinguishing between particles and antiparticles, etc.), or narrowing down to a particular subset of jets  (e.g., those containing a lepton), could be relevant to specific applications of fragmentation tagging. They can be addressed with the same general approach.

The rest of the paper is organized as follows. Section~\ref{sec: event simulation} describes the simulated event samples used in our study. In section~\ref{sec: basic variables}, we analyze a number of simple variables that can be used for classification. In section~\ref{sec: ML}, we present the neural network architectures we used (whose more detailed descriptions are provided in appendix~\ref{app: NN details}) and the resulting performance. Section~\ref{sec: summary and discussion} summarizes our results and conclusions.

\section{Event simulation}
\label{sec: event simulation}

To obtain representative samples of the jet types of interest, we proceed as follows.

\subsection{Event generation}
\label{subsec: event generation}

We simulate dijet events in proton-proton collisions with a center-of-mass energy of $14$~TeV. We use \textsc{MadGraph5} 3.5.1~\cite{alwall2014automated} to create the hard scattering process at the leading order and \textsc{Pythia} 8.308~\cite{bierlich2022comprehensive} to handle the parton showering, hadronization and decays. We define two kinematic regions for the jets (clustered as described below): $p_{T,{\rm jet}} > 200$~GeV and $p_{T,{\rm jet}} > 45$~GeV. These jet $p_T$ cuts follow generation-level cuts on the partons of $p_T > 180$~GeV and $p_T > 35$~GeV, respectively, and pseudorapidity $|\eta| < 4$.

To obtain samples of $s$ and $d$ jets, we generate $s\bar s$ and $d\bar d$ events. We include only the $gg$ and $u\bar u$ initial states so that the only difference between the $s$ and $d$ jets in the resulting samples will be their flavor.\footnote{If we included the $d\bar d$ initial state, for example, its contribution would be different for the $d\bar d$ and $s\bar s$ final states in terms of both its size (relative to the $gg$ contribution) and kinematics, due to the presence of a $t$-channel diagram in the first case only. Including the $s\bar s$ initial state as well would not compensate for this effect because of the lower $s$-quark content in the proton.} We do it to ensure that the taggers rely solely on the intrinsic differences between the jets and not on differences between their $p_T$ and $\eta$ distributions in particular samples.

To obtain samples of $b$ jets, we generate $b\bar{b}$ events from $gg$ and $q\bar q$ (with $q = u,d,s$) initial states. The $b$ jets are separated into a sample containing $b$ baryons and a sample containing $b$ mesons. The baryon samples are dominated by the $\Lambda_b^0$ (with smaller contributions from $\Xi_b^0$, $\Xi_b^-$, and $\Omega_b^-$), and the meson samples consist of $\overline B^0$ ($\sim 45\%$), $B^-$ ($\sim 45\%$), and $\overline B_s^0$ ($\sim 10\%$). Their antiparticles are included.

\subsection{Detector simulation}
\label{subsec: detector simulation}

We consider particles in the range $|\eta| < 4$, which is approximately the range that will be covered by the ATLAS and CMS tracking detectors at the HL-LHC~\cite{atlas2019expected,cms2018expected}. In addition, charged particles need to satisfy $p_T > 0.5$~GeV. Particles are treated as stable if they do not decay within $1$\,m from the beam axis and $3$\,m along the beam axis from the interaction point.\footnote{This roughly describes the inner detector volume in ATLAS~\cite{ATLAS:2008xda} and CMS~\cite{CMS:2008xjf}. Beyond this volume, particles are absorbed in the calorimeters.} Based on them, we form the following detector-level objects with the help of the Monte Carlo truth information:
\begin{itemize}

\item Charged hadrons.

We simulate track reconstruction efficiency as a function of the track production radius $r_{\rm prod}$ (relative to the beam axis) according to the expectations for the ATLAS tracker at the HL-LHC~\cite{strebler2022expected}. This efficiency starts at about $95\%$ for $r_{\rm prod} = 0$, decreases gradually to $65\%$ at $r_{\rm prod} \approx 38$~cm, and then drops sharply. For $r_{\rm prod} > 50$~cm, we set the reconstruction efficiency to zero. Charged hadrons that fail track reconstruction are counted as neutral hadrons.

For $b$ jets, we distinguish between charged hadrons originating from the primary vertex (PV), which are produced simultaneously with the $b$ hadron during hadronization, and those produced from the decay of the $b$ hadron, thus originating from a secondary vertex (SV). This distinction is achieved in our simulation based on truth information. If a charged hadron has a $b$ hadron as one of its ancestors, it is categorized as originating from the SV. Otherwise, it is classified as originating from the PV.

\item Neutral hadrons.

Associated with the typical HCAL granularity, we implement a grid with cell sizes of $0.1 \times 0.1$ in the $\eta$-$\phi$ space and calculate the energy contributed by the neutral hadrons to each cell within this grid. Each populated cell is then described as a single neutral hadron, regardless of how many neutral hadrons actually fell within its boundaries. While charged hadrons deposit their energy in the HCAL as well, it can be approximately subtracted based on the momentum measurement of their tracks in the tracker. Hence, in our simulation, we exclude the energy of charged hadrons from the HCAL measurements, except for those that fail track reconstruction.

\item Photons ($\gamma$).

Associated with the typical ECAL granularity, we implement a grid with cell sizes of $0.02 \times 0.02$ in the $\eta$-$\phi$ space. Photons contributing to the same cell are considered a single photon. We assume that contributions from electrons (except for those that fail track reconstruction) are subtracted based on their track measurements and neglect the electromagnetic energy depositions due to muons and charged hadrons.

\item Electrons ($e^\pm$), except for those failing track reconstruction and then counted as photons.

\item Muons ($\mu^\pm$), except for those failing track reconstruction.

\end{itemize}

\subsection{Reconstruction of \texorpdfstring{$K_S$ and $\Lambda$}{KS and Lambda} decays}
\label{subsec: KS Lambda reconstruction}

Since energetic $K_S$ mesons and $\Lambda$ baryons are more common in strange than in down-quark jets, it is useful for the purpose of strange-quark tagging to attempt identifying them from their decay products. In addition, since $\Lambda$ baryons are more common in $b$-baryon than in $b$-meson decays, while $K_S$ mesons are more common in $b$-meson decays, reconstructing $K_S$ and $\Lambda$ decays is also useful for the purpose of $b$-baryon/$b$-meson discrimination.

The $K_S$ meson decays as $K_S \to \pi^+\pi^-$ with a 69\% branching ratio, and $K_S \to \pi^0\pi^0$ with a 31\% branching ratio. The $\Lambda$ baryon decays as $\Lambda \to p\pi^-$ with a 64\% branching ratio, and $\Lambda \to n\pi^0$ with a 36\% branching ratio. We emulate the reconstruction of $K_S$ and $\Lambda$ as intermediate particles for decays to charged hadrons that occur at a distance greater than $0.5$\,cm and less than $50$\,cm from the beam axis. The lower bound helps to avoid confusion with prompt tracks that originate from the interaction point, while the upper bound represents the radius beyond which track reconstruction becomes essentially impossible due to an insufficient number of tracker layers that can be used. Within this volume, $K_S$ and $\Lambda$ decays can be easily identified using their highly displaced vertices and the possibility to reconstruct the invariant mass of the two products, as long as the tracks are reconstructed (see, e.g., refs.~\cite{ATLAS:2023nze,ATLAS:2024nbm,CMS:2013zgf}). As mentioned above, we simulate the track reconstruction efficiency as a function of the track production radius based on ref.~\cite{strebler2022expected}. If both tracks from the $K_S$ or $\Lambda$ decay are reconstructed, we remove them from the list of charged hadrons and store them as a reconstructed $K_S$ or $\Lambda$ object instead.

Apart from the $\Lambda$ baryon, other relatively long-lived hadrons that are common in $b$-baryon decays are the $\Sigma^+$ and $\Sigma^-$ baryons~\cite{ParticleDataGroup:2022pth,Gronau:2018vei}. The dominant decays of these particles, $\Sigma^+ \to p\pi^0$, $\Sigma^+ \to n\pi^+$, and $\Sigma^- \to n\pi^-$, produce kinked track signatures. Each of the two segments of the kinked track may or may not be reconstructible, depending on the tracker layers it passes through. Due to this nontrivial dependence on the specifics of the tracking detector and tracking algorithms, we will not consider the identification of these signatures in this paper, for simplicity. See, however, related studies in refs.~\cite{ATLAS:2017oal,ATLAS:2022rme,CMS:2019ybf,CMS:2020atg,CMS:2023mny,Kats:2023gul}.

\subsection{Jet clustering and preprocessing}
\label{subsec: jet clustering}

The objects defined in sections~\ref{subsec: detector simulation} and \ref{subsec: KS Lambda reconstruction} are clustered into jets using the anti-$k_t$ algorithm~\cite{Cacciari:2008gp,cacciari2012fastjet} with a radius parameter $R=0.4$.

We consider the two leading jets in each event. In the case of $s\bar{s}$ or $d\bar{d}$ production, we assume the two leading jets to be $s$-quark or $d$-quark jets, respectively. Quark and antiquark jets are included together in our samples, but we note that for some applications it can be useful to treat them separately. For $b\bar b$ production, we include $b$ hadrons as soft ghost particles during jet clustering, and then examine which jets contain a $b$ baryon and which ones contain a $b$ meson. Jets without $b$ hadrons are discarded. After this procedure, the ghost particles are removed from the jets.

Properties of the jets and their constituents are recorded for the analysis. Recorded jet properties are $p_T$, $\eta$, the number of constituents, and the fractions of the jet energy contributed by each type of constituent: photon energy $E_{\gamma}$, electron energy $E_e$, muon energy $E_{\mu}$, charged hadron energy $E_{\text{CH}}$, neutral hadron energy $E_{\text{NH}}$, reconstructed $K_S \to \pi^+\pi^-$ energy $E_{K_S}$, and reconstructed $\Lambda \to p\pi^-$ energy $E_{\Lambda}$ (including $\overline\Lambda$). In the case of $b$ jets, instead of $E_{\text{CH}}$, we use two separate variables, $E_{\text{CH,PV}}$ and $E_{\text{CH,SV}}$, for charged hadrons originating from the primary vertex and those from the secondary vertex, respectively, as detailed in section~\ref{subsec: detector simulation}.

For the jet constituents, the identities are recorded in binary form. We also record the transverse momentum $p_T$ of each constituent $i$, normalized with respect to the jet $p_T$, 
\begin{equation}
    p_{T,i}^{\rm norm} = \frac{p_{T,i}}{p_{T,{\rm jet}}} \;.
\end{equation}
The positions of the jet constituents in the $\eta$-$\phi$ space, $(\eta_i, \phi_i)$, are expressed in terms of polar coordinates $(r,\alpha)$ centered on the jet axis, such that
\begin{align}
    \eta_i - \eta_{\rm jet} = r_i\cos\alpha_i \,,\\
    \phi_i - \phi_{\rm jet} = r_i\sin\alpha_i \,,
\end{align}
where $\alpha = 0$ is defined by the location of the most energetic constituent. In addition, we flip the signs of all $\alpha_i$ values if the total momentum on the left side of the $\alpha = 0$ line is greater than on the right side. This standardizes the data and removes physical redundancies before feeding the data into the neural networks.

\section{Basic discriminating variables}
\label{sec: basic variables}

In this section, we look at the distributions of several simple variables that characterize the jets. Some of them could potentially be used to distinguish between $s$-quark and $d$-quark jets, or between $b$-baryon and $b$-meson jets.

\subsection{Strange vs. down jets}

\begin{figure}[t!]
    \centering
    \begin{subfigure}{0.49\textwidth}
        \includegraphics[width=\textwidth]{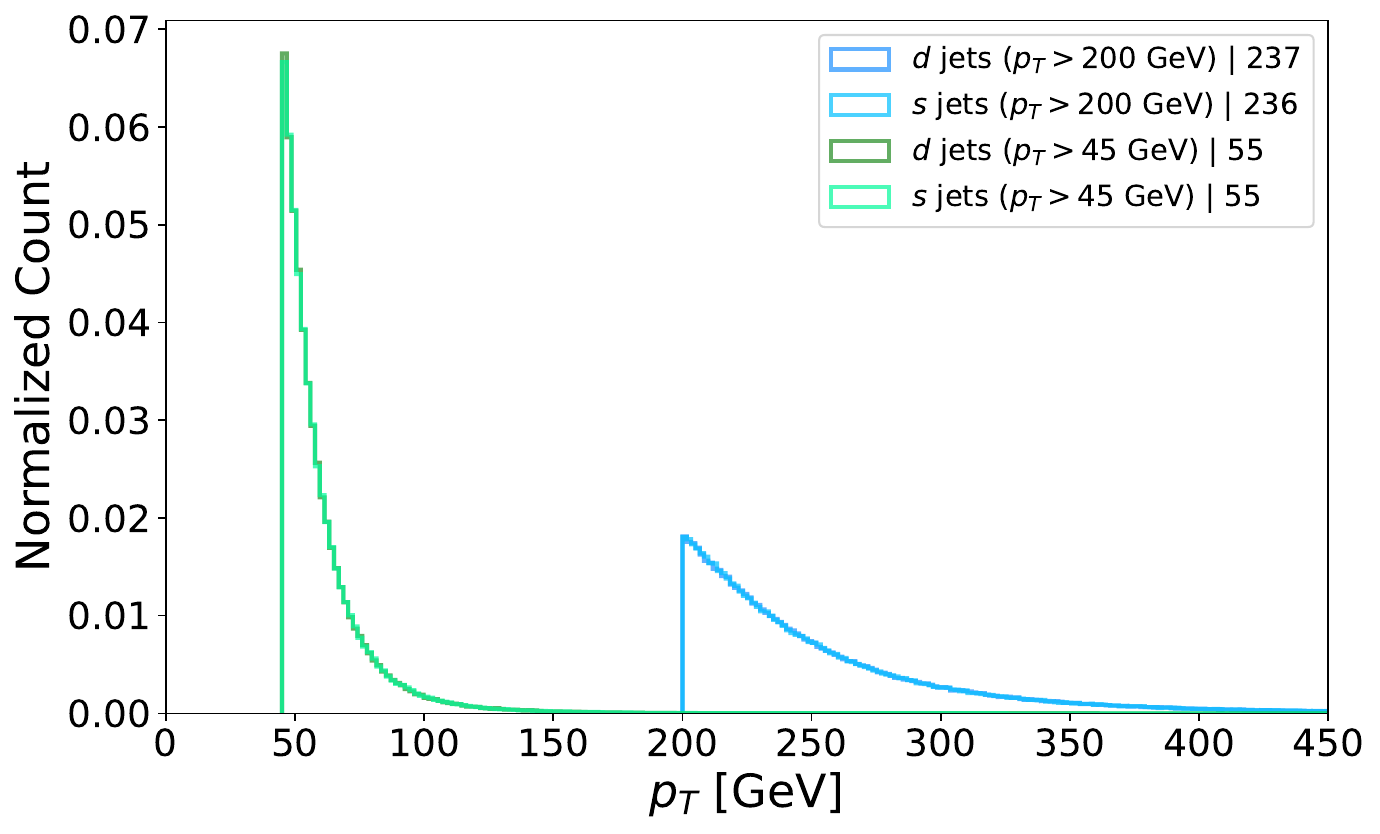}
        \caption{Jet transverse momentum distributions.}
        \label{fig:Jets_pT_Distribution}
    \end{subfigure}
    \hfill % horizontal fill command that inserts a blank space that will stretch accordingly to fill the space available
    \begin{subfigure}{0.49\textwidth}
        \includegraphics[width=\textwidth]{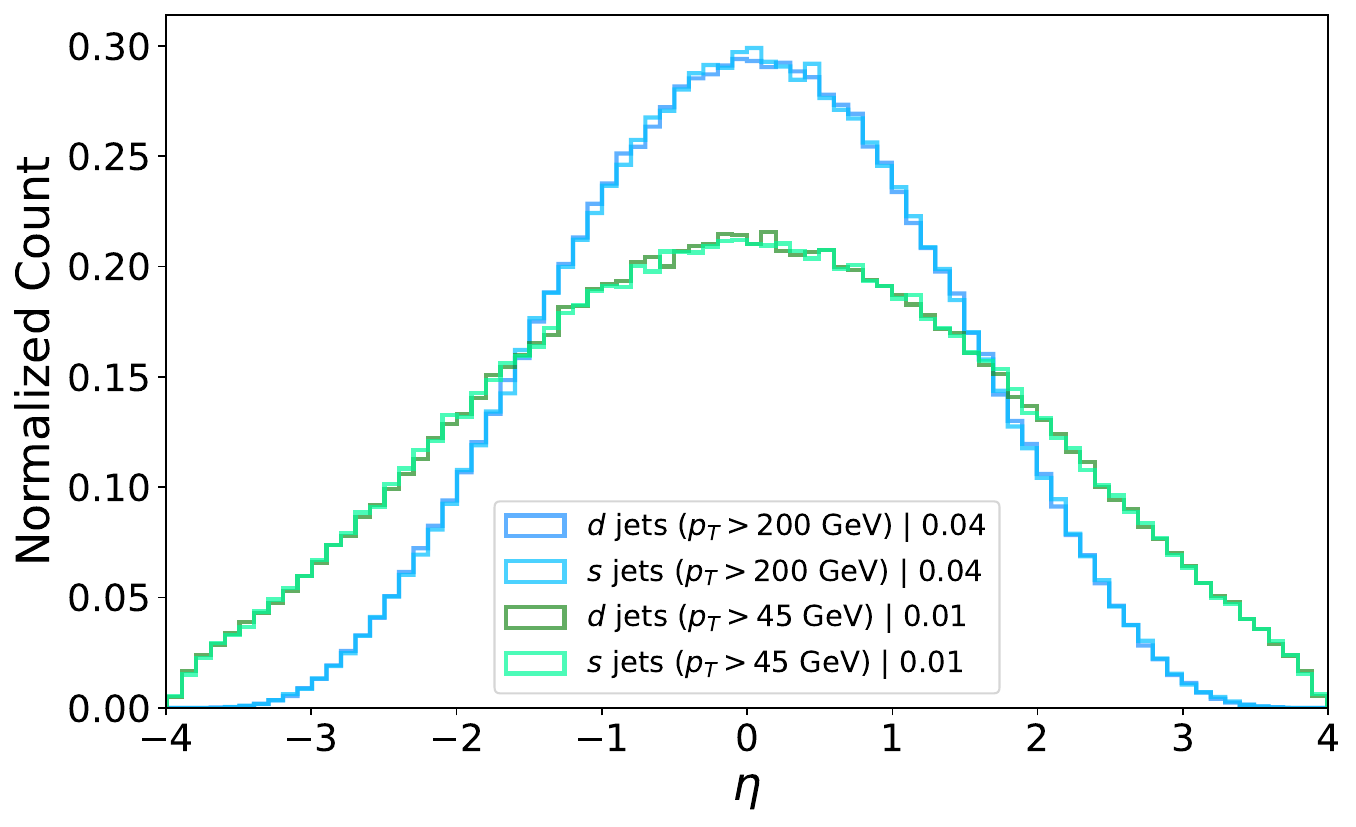}
        \caption{Jet pseudorapidity distributions.}
        \label{fig:Jet_eta_Distribution}
    \end{subfigure}
    \vskip 0.6\baselineskip %inserts a vertical space equivalent to the normal line spacing, effectively creating a blank line
    \begin{subfigure}{0.49\textwidth}
        \includegraphics[width=\textwidth]{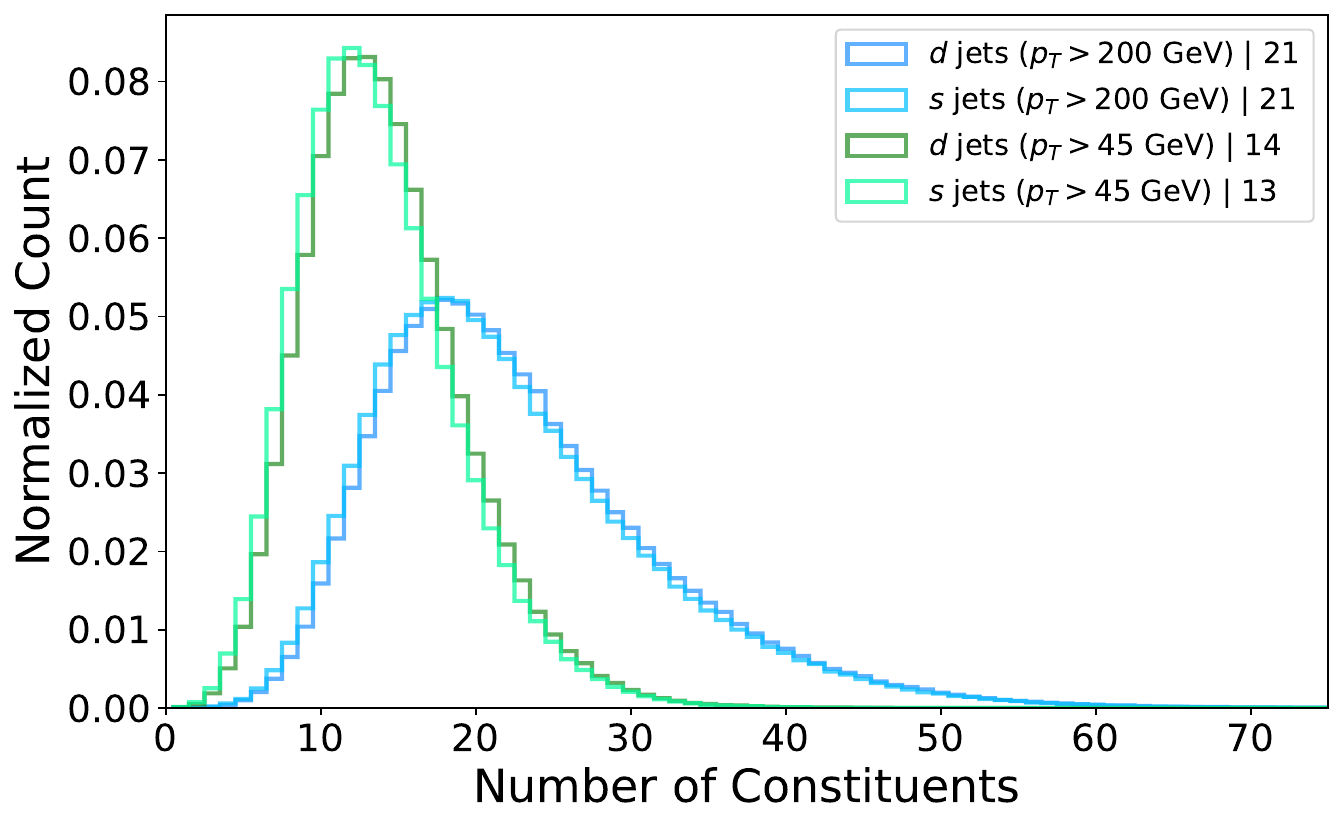}
        \caption{Number of constituents distributions.}
        \label{fig:Jet_Constituents_Number_Distribution}
    \end{subfigure}    
    \caption{Properties of $d$-quark and $s$-quark jets in our samples for $p_{T,{\rm jet}} > 200$~GeV and $p_{T,{\rm jet}} > 45$~GeV. Median values are given in the legends.}
    \label{fig:Jet_distributions}
\end{figure}

We first examine variables that characterize the entire jet, including $p_{T,{\rm jet}}$, $\eta_{\rm jet}$, and the number of constituents, for $s$-quark and $d$-quark jets. Their distributions are shown in figures~\ref{fig:Jets_pT_Distribution}, \ref{fig:Jet_eta_Distribution}, and~\ref{fig:Jet_Constituents_Number_Distribution}, respectively. The distributions of $p_{T,{\rm jet}}$ and ${\eta_{\rm jet}}$ are essentially identical between the $d$-quark and $s$-quark jets in our samples, as expected from our choice of the production processes. The number of constituents is higher for higher-$p_T$ jets, as expected, and tends to be slightly higher in $d$-quark jets than in $s$-quark jets.

\begin{figure}
    \centering
    \vskip\baselineskip
    \begin{subfigure}{0.49\textwidth}
        \includegraphics[width=\textwidth]{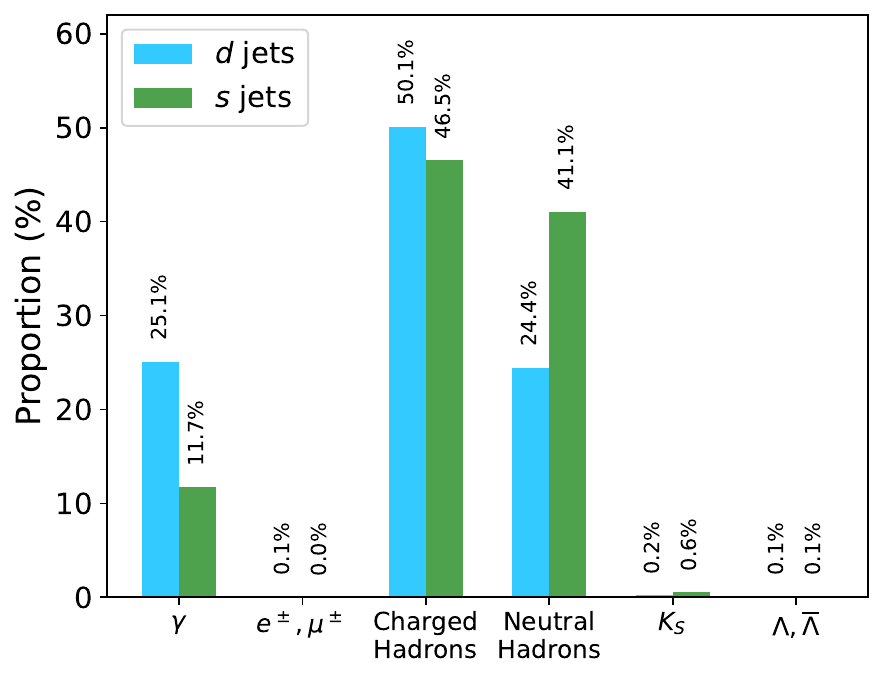}
        \caption{$p_{T,{\rm jet}} > 200$~GeV}
    \end{subfigure}
    \begin{subfigure}{0.49\textwidth}
        \includegraphics[width=\textwidth]{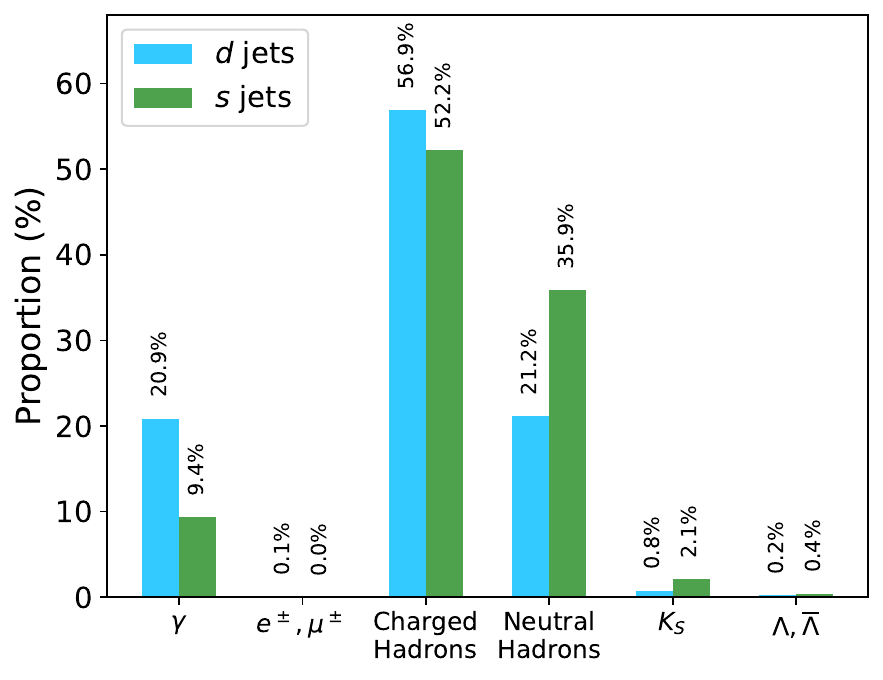}
        \caption{$p_{T,{\rm jet}} > 45$~GeV}
    \end{subfigure}
    \caption{Distributions of the constituent types with the highest transverse momentum ($p_T$) in $s$-quark and $d$-quark jets for (a) $p_{T,{\rm jet}} > 200$~GeV, (b) $p_{T,{\rm jet}} > 45$~GeV.}
    \label{fig: Highest_pT_particle_Count_Histogram}
\end{figure}

Figure~\ref{fig: Highest_pT_particle_Count_Histogram} shows the distributions of the identities of the constituents with the highest $p_T$ within each jet. As expected, we observe that it is more common for $s$-quark jets than for $d$-quark jets to have a neutral hadron or a reconstructed $K_S$ as the most energetic constituent, while it is the other way around for photons. It is related to the fact that an $s$ quark often produces an energetic $K_L$ or $K_S$ meson, while $d$ quarks produce energetic $\pi^0$ mesons (which decay to photons) more frequently.

\begin{figure}
    \centering
    \begin{subfigure}[b]{\textwidth}
        \includegraphics[width=\textwidth]{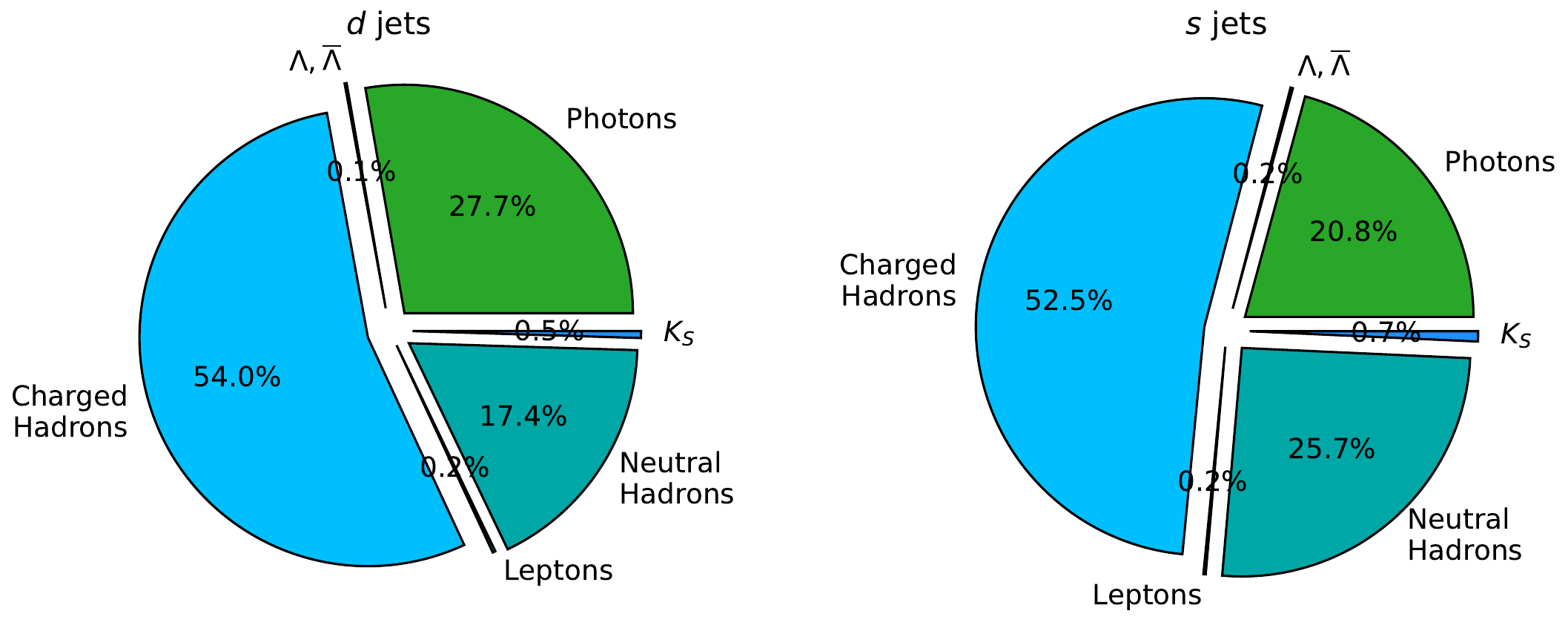}
        \caption{$p_{T,{\rm jet}} > 200$~GeV}
        \label{subfig: energy_pie_chart_200}
    \end{subfigure}
    \vskip\baselineskip %inserts a vertical space equivalent to the normal line spacing, effectively creating a blank line
    \begin{subfigure}[b]{\textwidth}
        \includegraphics[width=\textwidth]{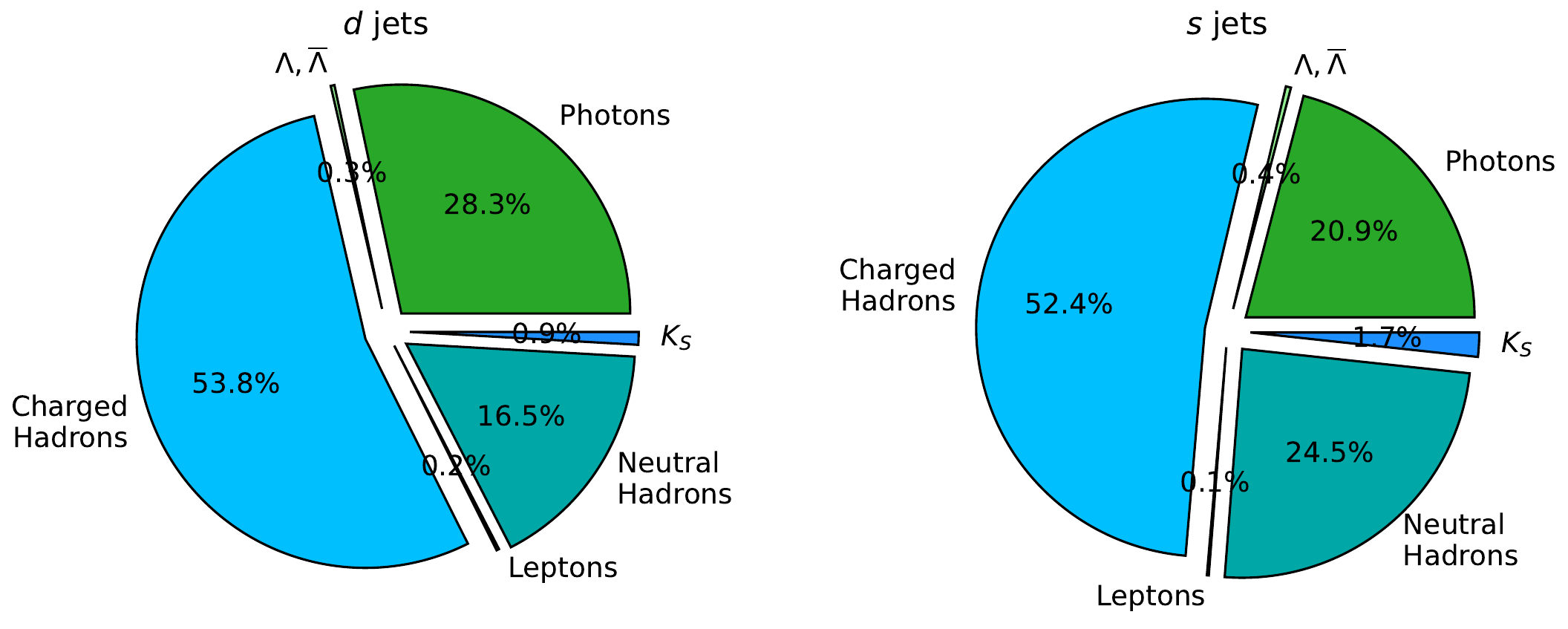}
        \caption{$p_{T,{\rm jet}} > 45$~GeV}
        \label{subfig: energy_pie_chart_45}
    \end{subfigure}
    \caption{The mean energy fractions contributed by the various types of constituents in $d$-quark jets (left) and $s$-quark jets (right) for (a) $p_{T,{\rm jet}} > 200$~GeV, (b) $p_{T,{\rm jet}} > 45$~GeV.}
    \label{fig: energy_pie_chart}
\end{figure}

Figure~\ref{fig: energy_pie_chart} shows the mean energy fraction contributed by each type of constituent within the jet. Related to the previous observations, we see that the energy in neutral hadrons (as well as reconstructed $K_S$ decays, especially for low-$p_T$ jets) is greater for $s$-quark jets, while the energy in photons is higher for $d$-quark jets. Consequently, $s$-quark jets deposit a larger proportion of their energy in the HCAL, while $d$-quark jets tend to deposit a greater fraction of their energy in the ECAL.

\begin{figure}
    \centering
    \begin{subfigure}{0.49\textwidth}
        \includegraphics[width=\textwidth]{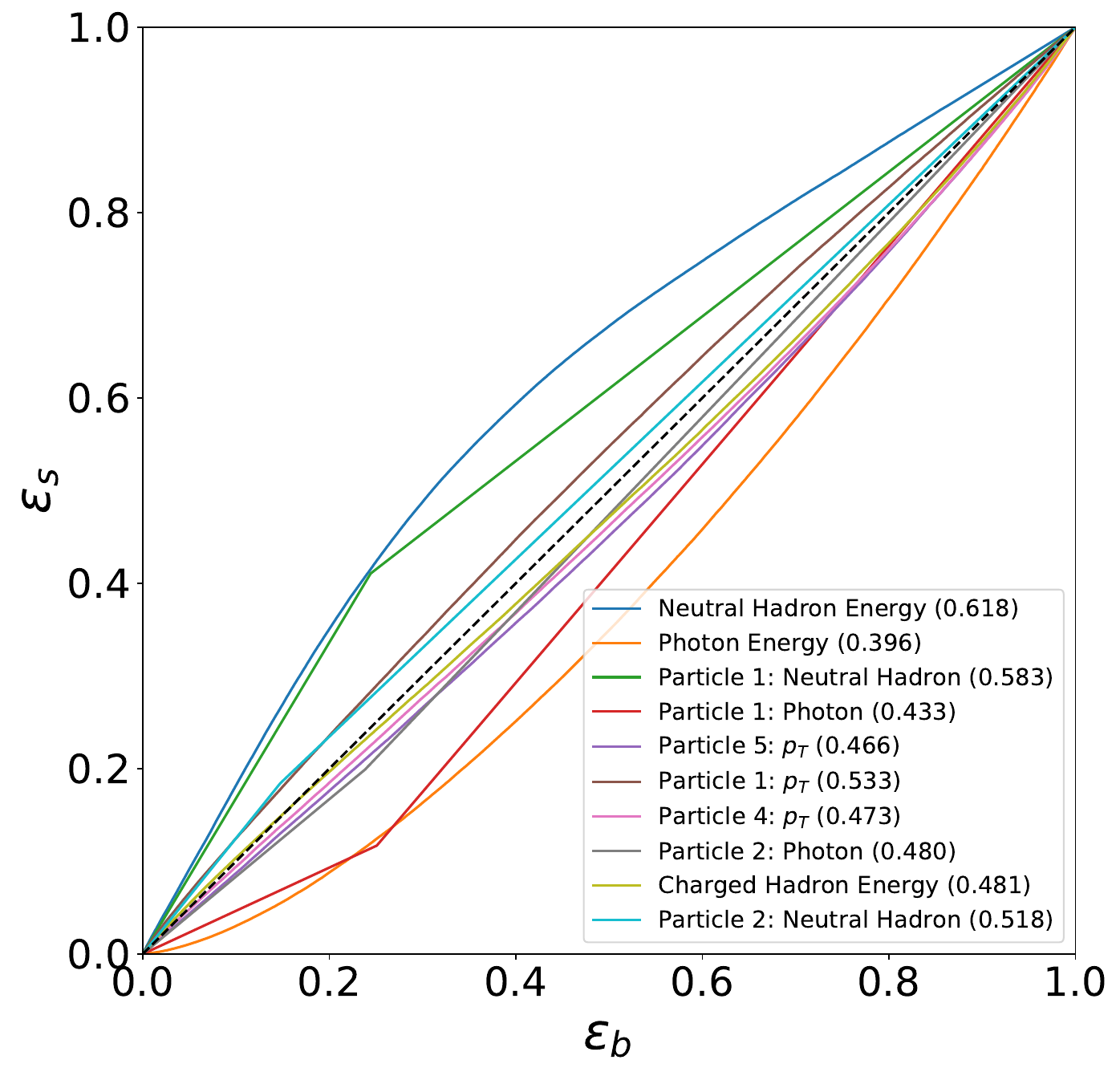}
        \caption{$p_{T,{\rm jet}} > 200$~GeV}
    \end{subfigure}
    %\vskip\baselineskip %inserts a vertical space equivalent to the normal line spacing, effectively creating a blank line
    \begin{subfigure}{0.49\textwidth}
        \includegraphics[width=\textwidth]{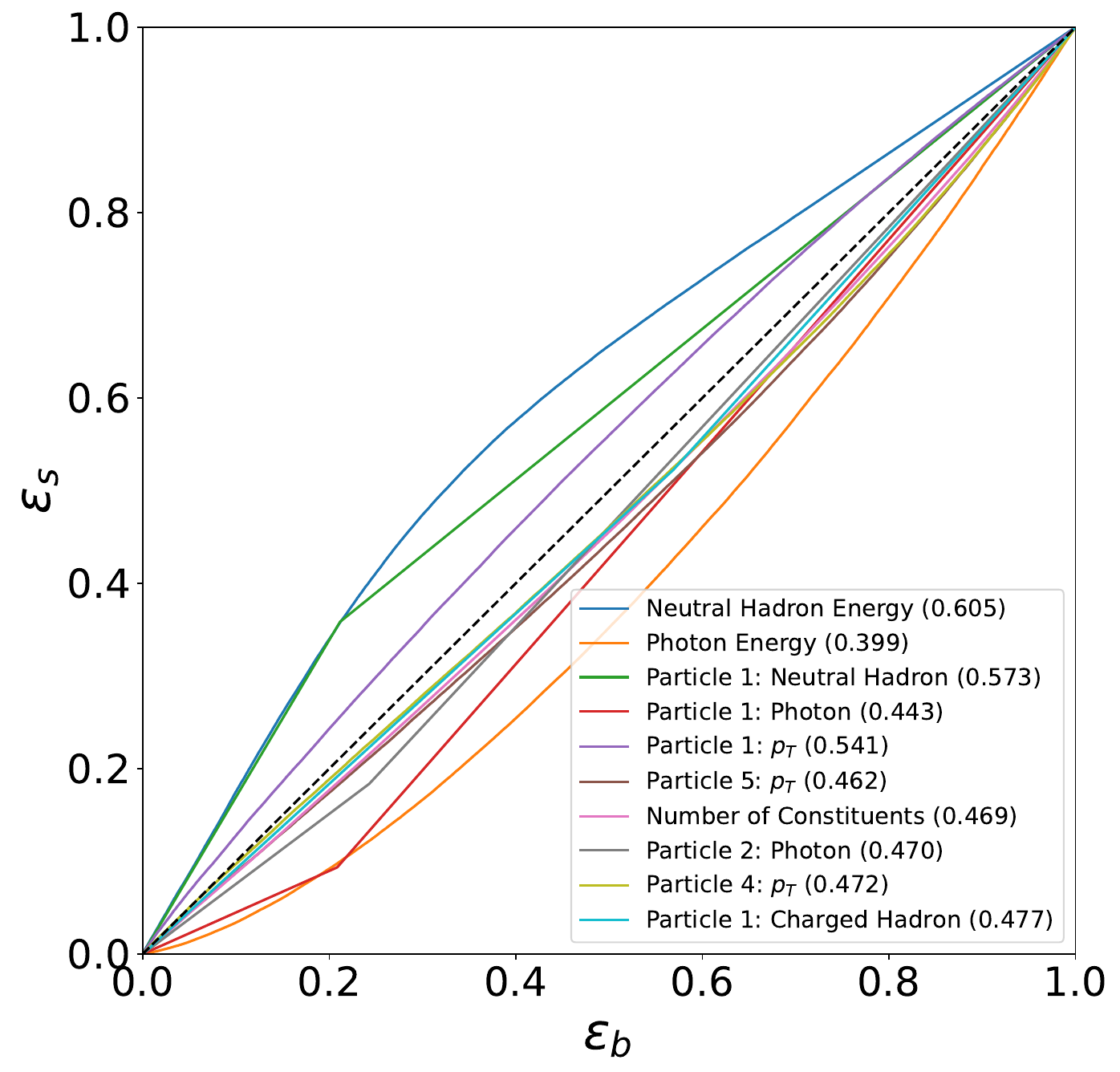}
        \caption{$p_{T,{\rm jet}} > 45$~GeV}
    \end{subfigure}
    \caption{ROC curves for the ten most discriminative features in $s$-quark (signal) and $d$-quark (background) jets for (a) $p_{T,{\rm jet}} > 200$~GeV and (b) $p_{T,{\rm jet}} > 45$~GeV. Both particle and jet-level features from section~\ref{subsec: jet clustering} are included. The particles are numbered by decreasing $p_T$ values. The kinks present in curves corresponding to particle identities are due to the binary nature of the variable. The AUC values are given in parentheses in the legends, where the features are ordered based on the absolute distance of their AUC from $0.5$.}
    \label{fig: ROC For Most Discriminative Jet Features}
\end{figure}

\begin{figure}[ht]
    \centering
    \begin{subfigure}[b]{1.0\textwidth}
        \includegraphics[width=\textwidth]{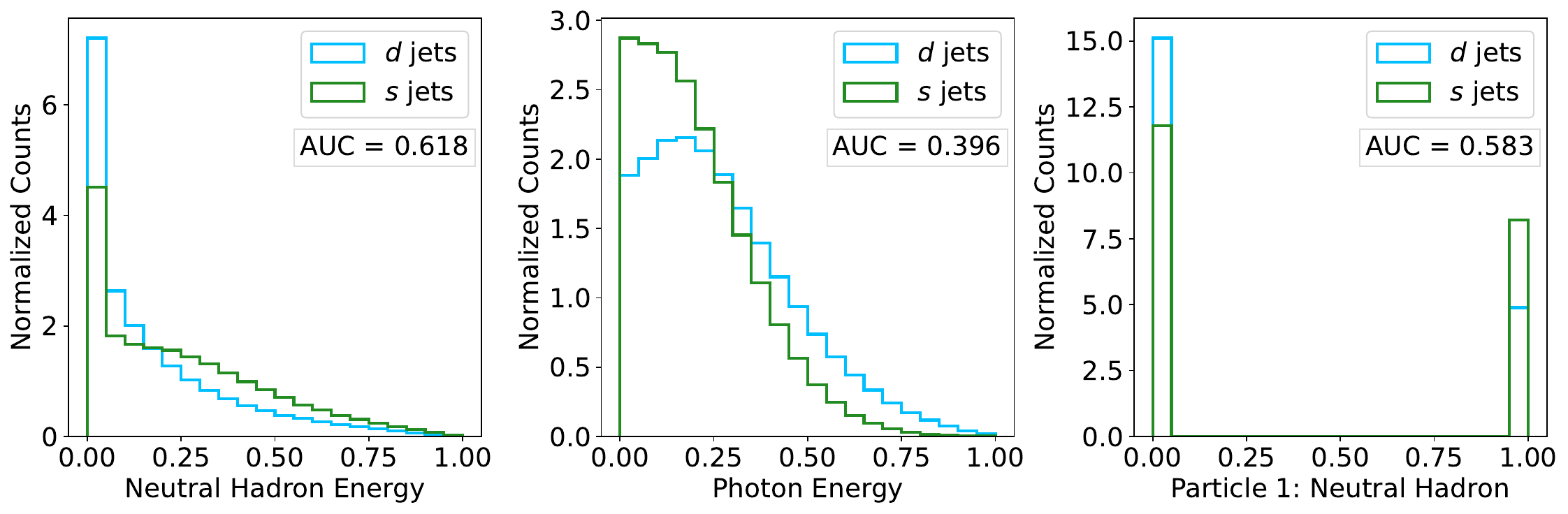}
        \caption{$p_{T,{\rm jet}} > 200$~GeV}
        \label{subfig: Distribution_3_Discriminative_Jet_Features_200}
    \end{subfigure}
    \vskip\baselineskip %inserts a vertical space equivalent to the normal line spacing, effectively creating a blank line
    \begin{subfigure}[b]{1.0\textwidth}
        \includegraphics[width=\textwidth]{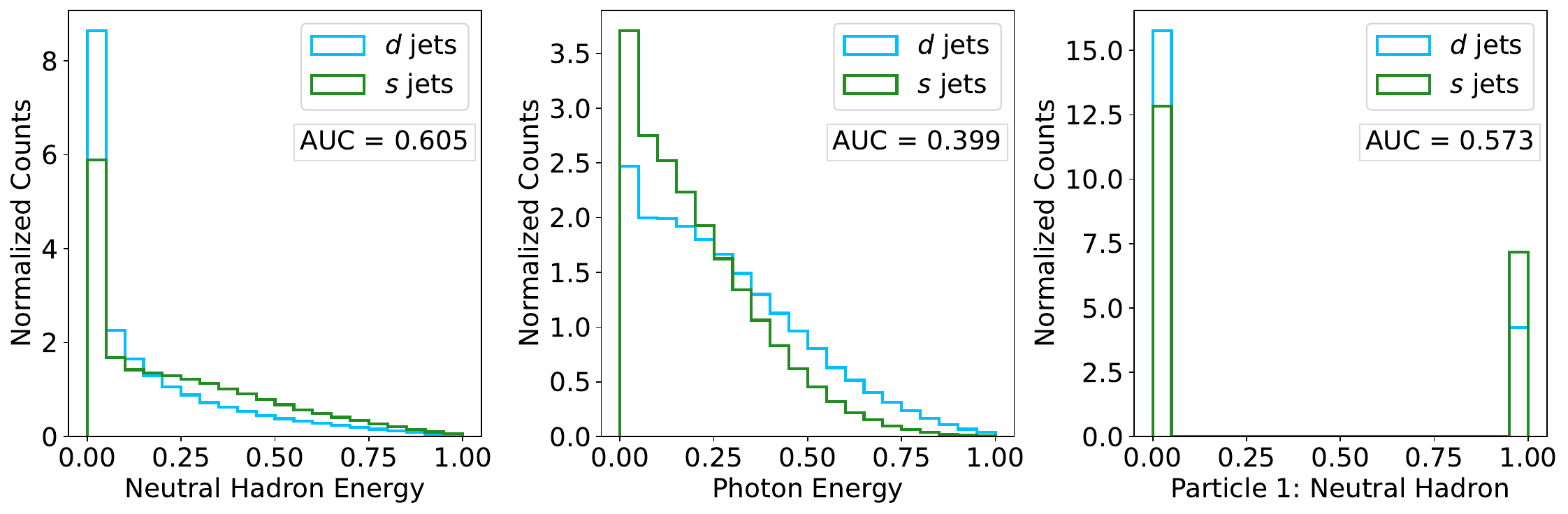}
        \caption{$p_{T,{\rm jet}} > 45$~GeV}
        \label{subfig: Distribution_3_Discriminative_Jet_Features_45}
    \end{subfigure}
    \caption{Distributions of the three most discriminative features from figure~\ref{fig: ROC For Most Discriminative Jet Features} in $d$-quark and $s$-quark jets for (a) $p_{T,{\rm jet}} > 200$~GeV, (b) $p_{T,{\rm jet}} > 45$~GeV.}
    \label{fig: s_d_Distribution_3_Discriminative_Jet_Features}
\end{figure}

To see to what extent the jet classification task can be accomplished by simply applying a cut on the distribution of any individual variable, we utilize Receiver Operating Characteristic (ROC) curves. An ROC curve characterizes the discriminating power of applying a threshold to a given variable. It presents the signal efficiency ($\varepsilon_s$) vs.\ the background efficiency ($\varepsilon_b$) achieved at varying threshold values. In the present case, the signal efficiency corresponds to correctly identifying $s$-quark jets, and the background efficiency indicates the fraction of $d$-quark jets incorrectly identified as $s$-quark jets. For each of the variables, we construct an ROC curve and compute the Area Under the Curve (AUC). The AUC serves as a metric for the discriminative power of the variable. Random guessing would give an AUC of $0.5$, whereas an AUC of $1$ indicates perfect discrimination, and an AUC of $0$ also signifies perfect discrimination but with an opposite threshold direction. Figure~\ref{fig: ROC For Most Discriminative Jet Features} presents the ROC curves for the ten most discriminative jet and constituent features.\footnote{Features of constituents beyond the fifth most energetic one are not considered for this plot since they are less likely to be meaningful as individual variables and because they are not available in all jets. However, features of all the constituents of each jet will be made available to the advanced neural networks.} The neutral hadron energy and photon energy, followed by the identity of the most energetic particle, whether it is a neutral hadron or a photon, show the highest/lowest AUC values, implying they are the most discriminative features. Figure~\ref{fig: s_d_Distribution_3_Discriminative_Jet_Features} shows the distributions of the three most discriminative features identified in figure~\ref{fig: ROC For Most Discriminative Jet Features}.

\FloatBarrier

\subsection{Bottom baryon vs. meson jets}

In this section, we extend our analysis to distinguishing between $b$-baryon and $b$-meson jets, which is an example of fragmentation tagging.

\begin{figure}[t!]
    \centering
    \begin{subfigure}{0.49\textwidth}
        \includegraphics[width=\textwidth]{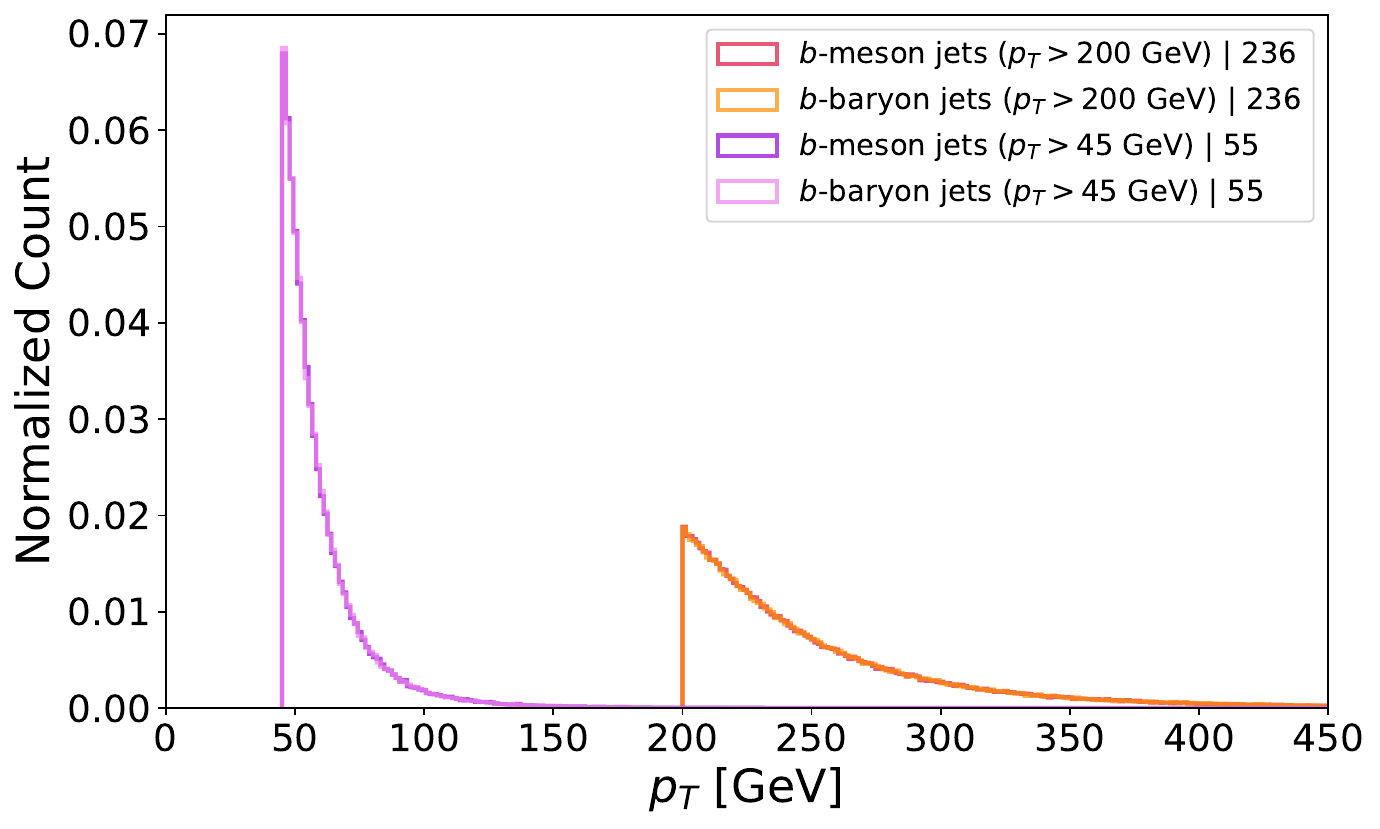}
        \caption{Jet transverse momentum distributions.}
        \label{fig:b_Jets_pT_Distribution}
    \end{subfigure}
    \hfill % horizontal fill command that inserts a blank space that will stretch accordingly to fill the space available
    \begin{subfigure}{0.49\textwidth}
        \includegraphics[width=\textwidth]{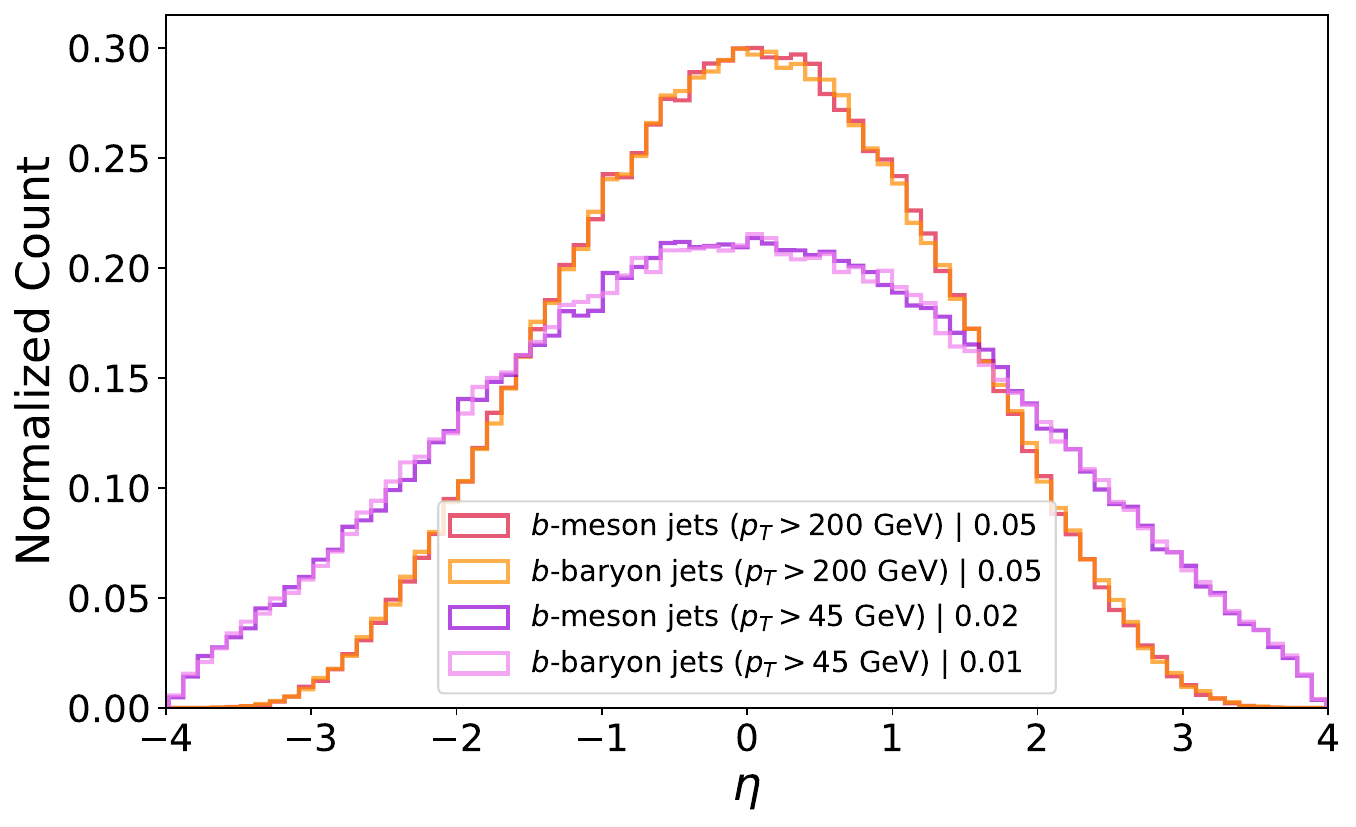}
        \caption{Jet pseudorapidity distributions.}
        \label{fig:b_Jet_eta_Distribution}
    \end{subfigure}
    \vskip 0.6\baselineskip %inserts a vertical space equivalent to the normal line spacing, effectively creating a blank line
    \begin{subfigure}{0.49\textwidth}
        \includegraphics[width=\textwidth]{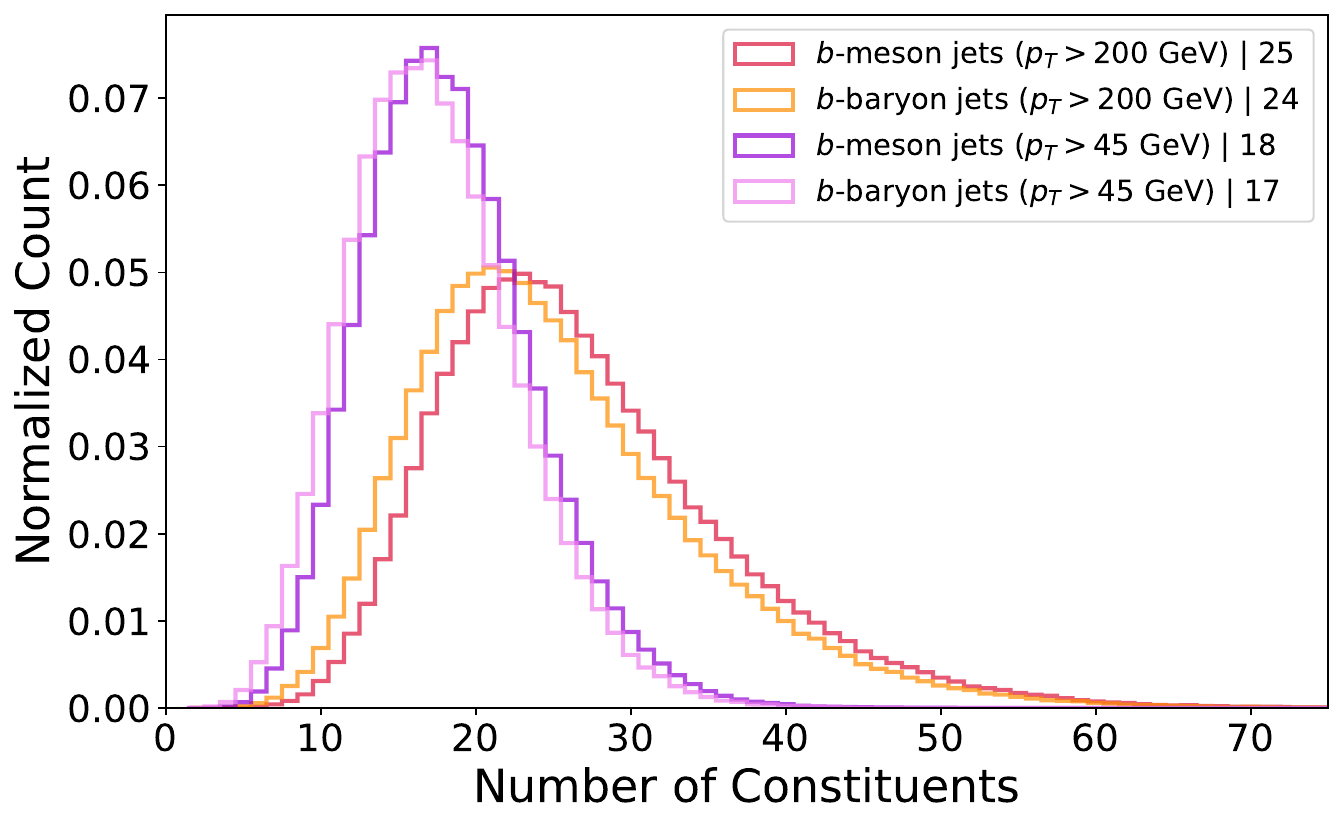}
        \caption{Number of constituents distributions.}
        \label{fig:b_Jet_Constituents_Number_Distribution}
    \end{subfigure}    
    \caption{Properties of $b$-meson and $b$-baryon jets in our samples for $p_{T,{\rm jet}} > 200$~GeV and $p_{T,{\rm jet}} > 45$~GeV. Median values are given in the legends.}
    \label{fig:b_Jet_distributions}
\end{figure}

Figure~\ref{fig:b_Jet_distributions} presents the distributions of $p_{T,{\rm jet}}$, $\eta_{\rm jet}$, and the number of constituents within $b$-meson and $b$-baryon jets. The distributions of $p_{T,{\rm jet}}$ and $\eta_{\rm jet}$ are essentially identical between the $b$-meson and $b$-baryon jets, as expected. We also see that jets containing $b$ mesons demonstrate a slightly higher constituent count, on average, than those containing $b$ baryons.

\begin{figure}
    \centering
    \vskip\baselineskip
    \begin{subfigure}{0.49\textwidth}
        \includegraphics[width=\textwidth]{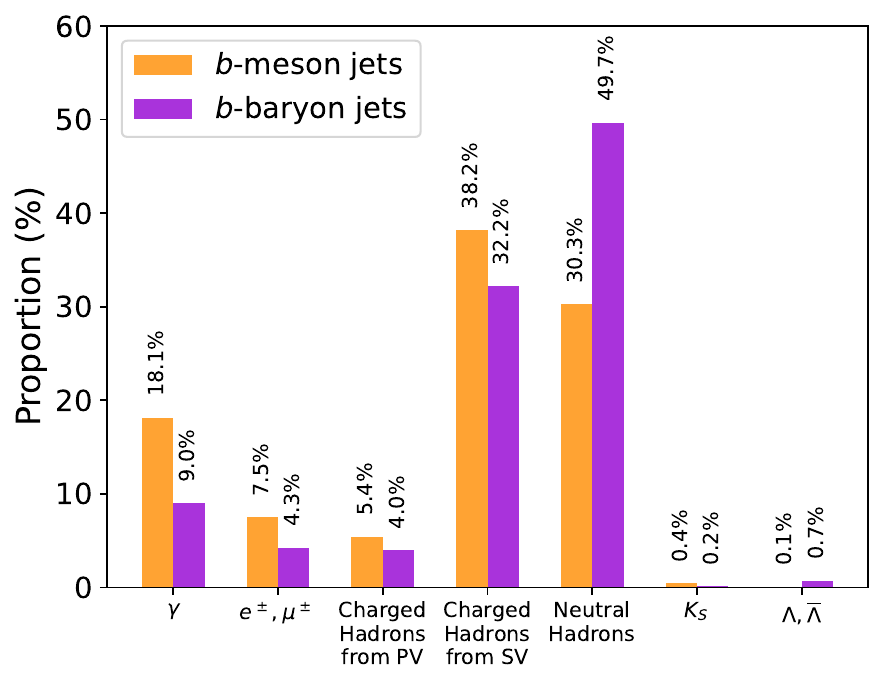}
        \caption{$p_{T,{\rm jet}} > 200$~GeV}
    \end{subfigure}
    \begin{subfigure}{0.49\textwidth}
        \includegraphics[width=\textwidth]{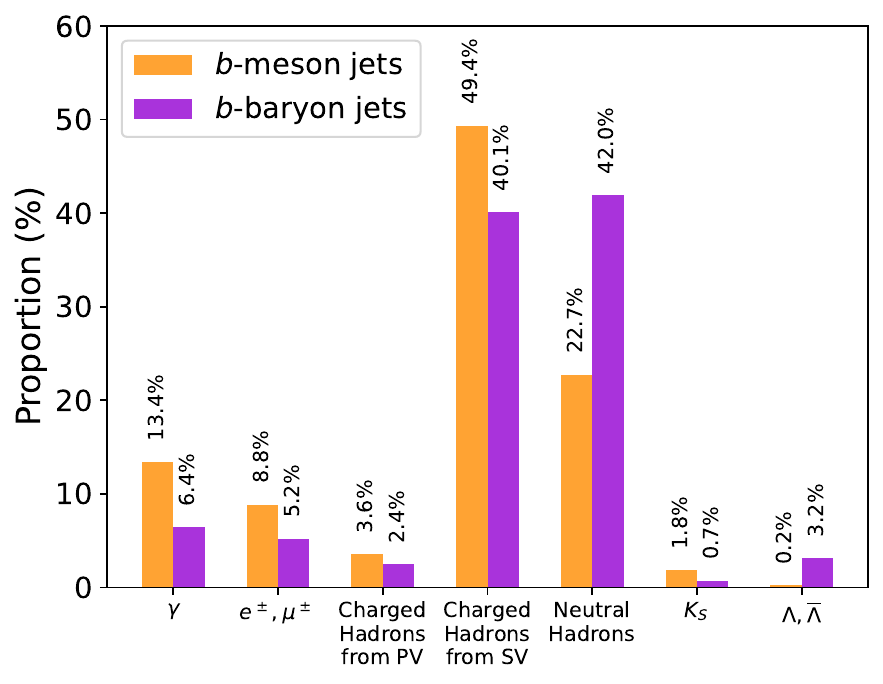}
        \caption{$p_{T,{\rm jet}} > 45$~GeV}
    \end{subfigure}
    \caption{Distributions of the constituent types with the highest transverse momentum ($p_T$) in $b$-meson and $b$-baryon jets for (a) $p_{T,{\rm jet}} > 200$~GeV, (b) $p_{T,{\rm jet}} > 45$~GeV.}
    \label{fig: b_Highest_pT_particle_Count_Histogram}
\end{figure}

Figure~\ref{fig: b_Highest_pT_particle_Count_Histogram} shows the distributions of the identities of the constituents with the highest $p_T$ in $b$ jets containing a $b$ baryon vs.\ those with a $b$ meson. We see that neutral hadrons are more common as the leading constituents in $b$-baryon jets. This can be attributed to the $\Lambda$ baryons and neutrons that are produced in many of the $b$-baryon decays, in line with baryon number conservation. While $b$-meson decays often produce neutral kaons, they will usually carry less energy due to their lower mass. Nevertheless, we see that reconstructed $K_S$ decays appear more frequently as the leading constituents in $b$-meson jets, and reconstructed $\Lambda$ decays in $b$-baryon jets, as expected.

Figure~\ref{fig: b_Highest_pT_particle_Count_Histogram} also shows that photons are more common as the leading constituents in $b$-meson jets. This can be attributed to $\pi^0$ decays. While pions are common in decays of all $b$ hadrons, the fact that $b$-baryon decay products necessarily include a baryon with a mass of about $1$~GeV leaves less room for energetic pions. Leptons are also more common in $b$-meson jets. While leptons from $b \to c$ transitions are expected to contribute similarly to both types of jets, leptons from $c \to s$ transitions are less common in $b$-baryon decays due to the small leptonic branching ratio of the $\Lambda_c^+$ (about $4\%$ per lepton flavor) relative to those of the $D$ mesons ($16\%$, $7\%$, and $6\%$ per flavor for the $D^+$, $D^0$, and $D_s^+$, respectively)~\cite{ParticleDataGroup:2022pth}. They are also less energetic because the necessity of having a baryon in the final state leaves less energy available to leptons.

\begin{figure}
    \centering
    \begin{subfigure}[b]{\textwidth}
        \includegraphics[width=\textwidth]{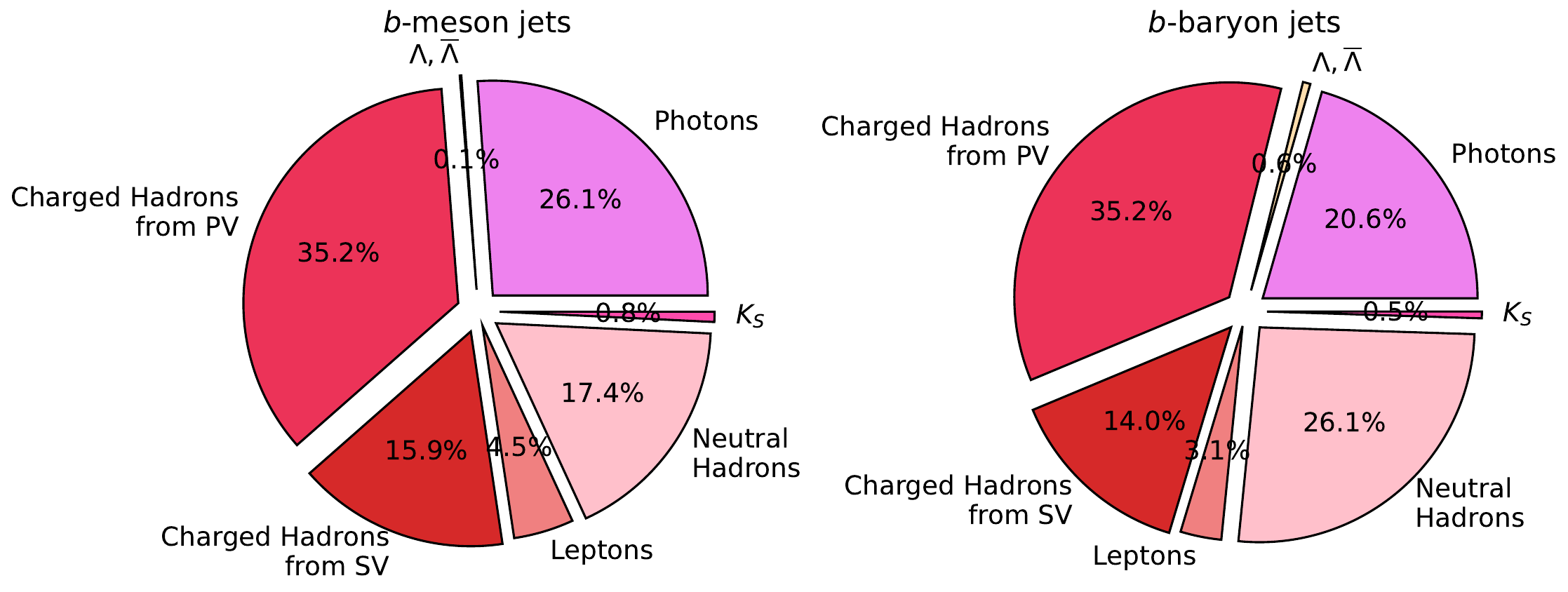}
        \caption{$p_{T,{\rm jet}} > 200$~GeV}
        \label{subfig: b_energy_pie_chart_200}
    \end{subfigure}
    \vskip\baselineskip %inserts a vertical space equivalent to the normal line spacing, effectively creating a blank line
    \begin{subfigure}[b]{\textwidth}
        \includegraphics[width=\textwidth]{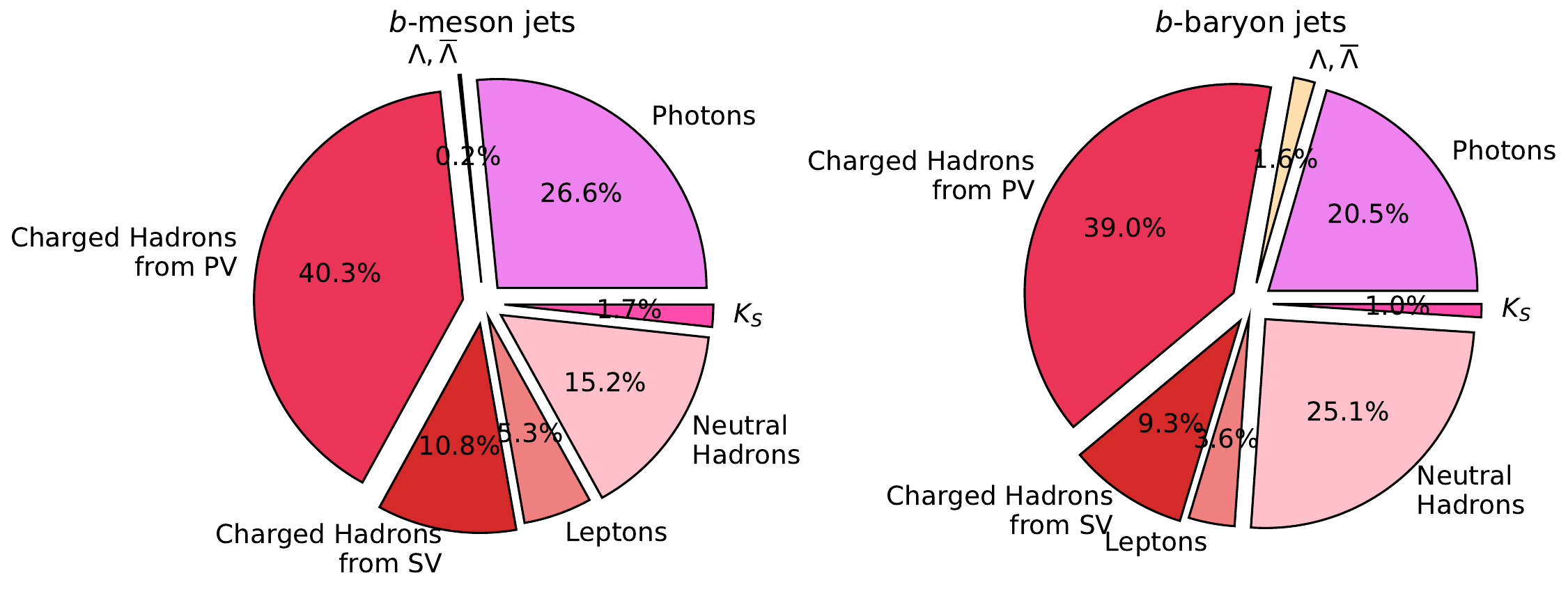}
        \caption{$p_{T,{\rm jet}} > 45$~GeV}
        \label{subfig: b_energy_pie_chart_45}
    \end{subfigure}
    \caption{The mean energy fractions contributed by the various types of constituents in $b$-meson jets (left) and $b$-baryon jets (right) for (a) $p_{T,{\rm jet}} > 200$~GeV, (b) $p_{T,{\rm jet}} > 45$~GeV. Leptons include electrons and muons.}
    \label{fig: b_energy_pie_chart}
\end{figure}

Figure~\ref{fig: b_energy_pie_chart} presents the mean energy fractions of each type of constituent within the jet. The behavior is similar to that observed for the most energetic constituent: the neutral hadronic energy and reconstructed $\Lambda$ energy are higher in $b$-baryon jets, while the energy fractions in photons, leptons, and reconstructed $K_S$ mesons are larger in $b$-meson jets.

\begin{figure}
    \centering
    \begin{subfigure}{0.49\textwidth}
        \includegraphics[width=\textwidth]{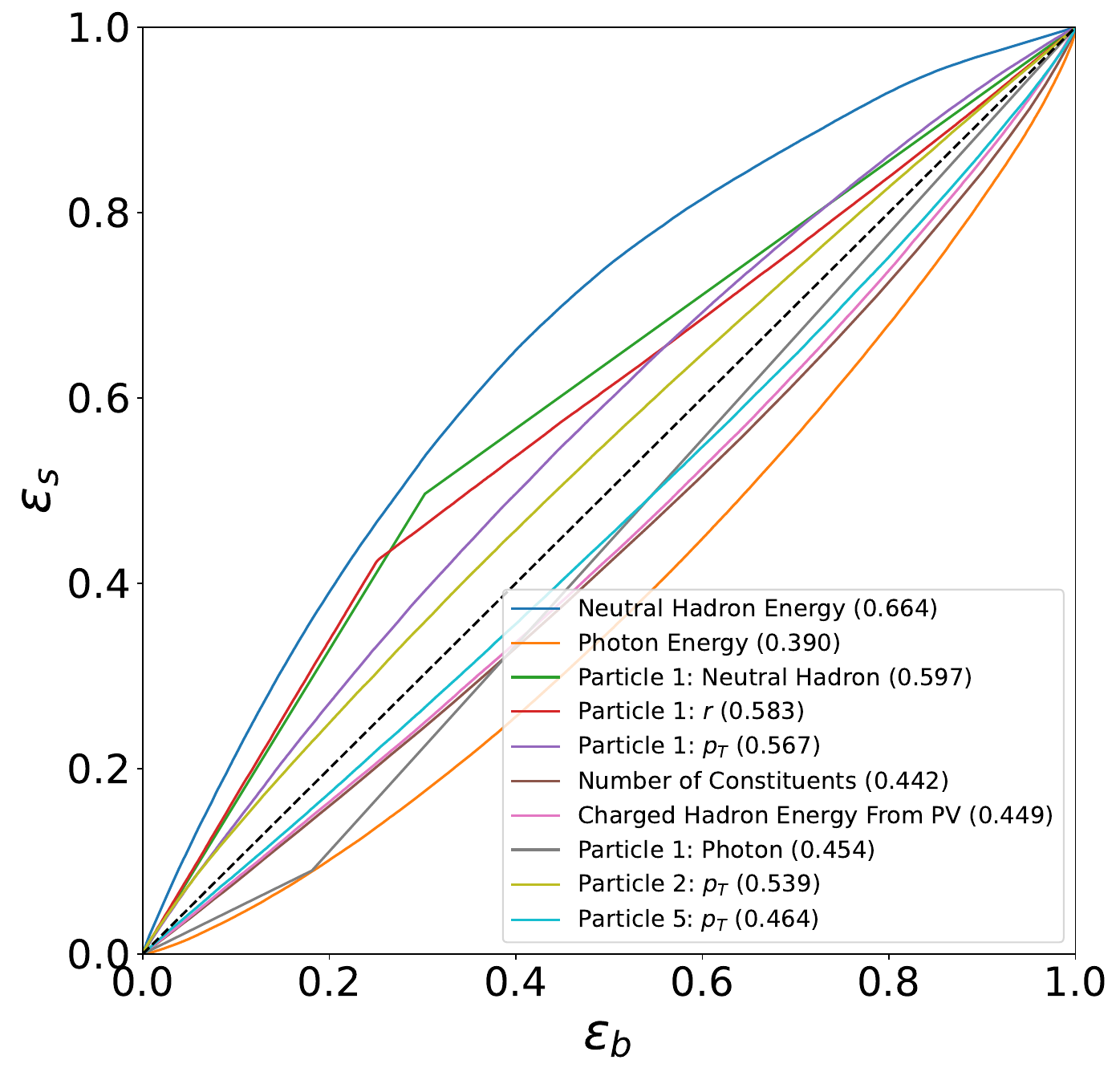}
        \caption{$p_{T,{\rm jet}} > 200$~GeV}
    \end{subfigure}
    %\vskip\baselineskip %inserts a vertical space equivalent to the normal line spacing, effectively creating a blank line
    \begin{subfigure}{0.49\textwidth}
        \includegraphics[width=\textwidth]{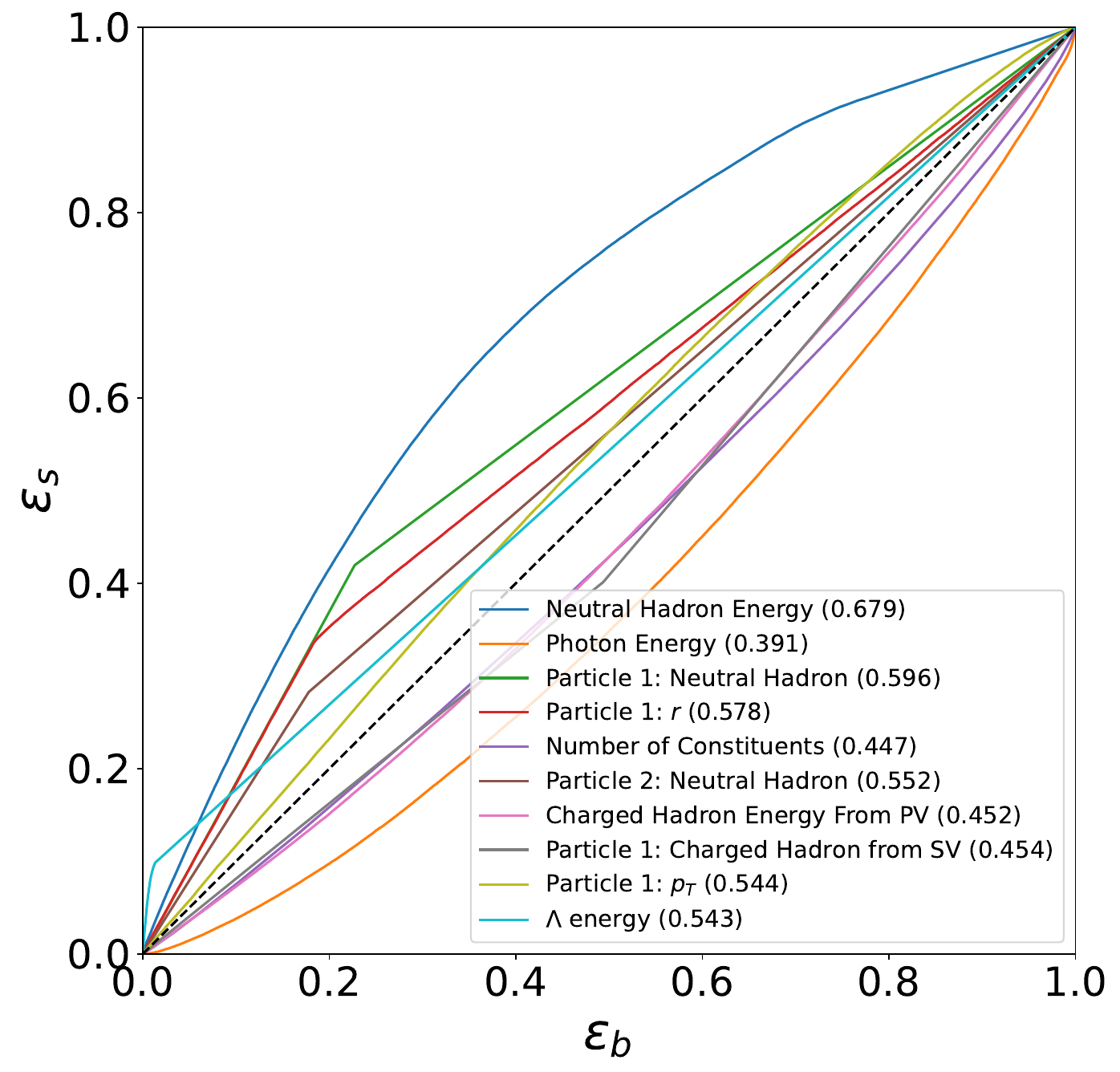}
        \caption{$p_{T,{\rm jet}} > 45$~GeV}
        \label{fig: b_ROC-45 For Most Discriminative Jet Features}
    \end{subfigure}
    \caption{ROC curves for the ten most discriminative features in $b$-baryon (signal) and $b$-meson (background) jets for (a) $p_{T,{\rm jet}} > 200$~GeV and (b) $p_{T,{\rm jet}} > 45$~GeV. Both particle and jet-level features from section~\ref{subsec: jet clustering} are included. The particles are numbered by decreasing $p_T$ values. The kinks present in curves corresponding to particle identities are due to the binary nature of the variable. The AUC values are given in parentheses in the legends, where the features are ordered based on the absolute distance of their AUC from $0.5$.}
    \label{fig: b_ROC For Most Discriminative Jet Features}
\end{figure}

\begin{figure}[ht]
    \centering
    \begin{subfigure}[b]{1.0\textwidth}
        \includegraphics[width=\textwidth]{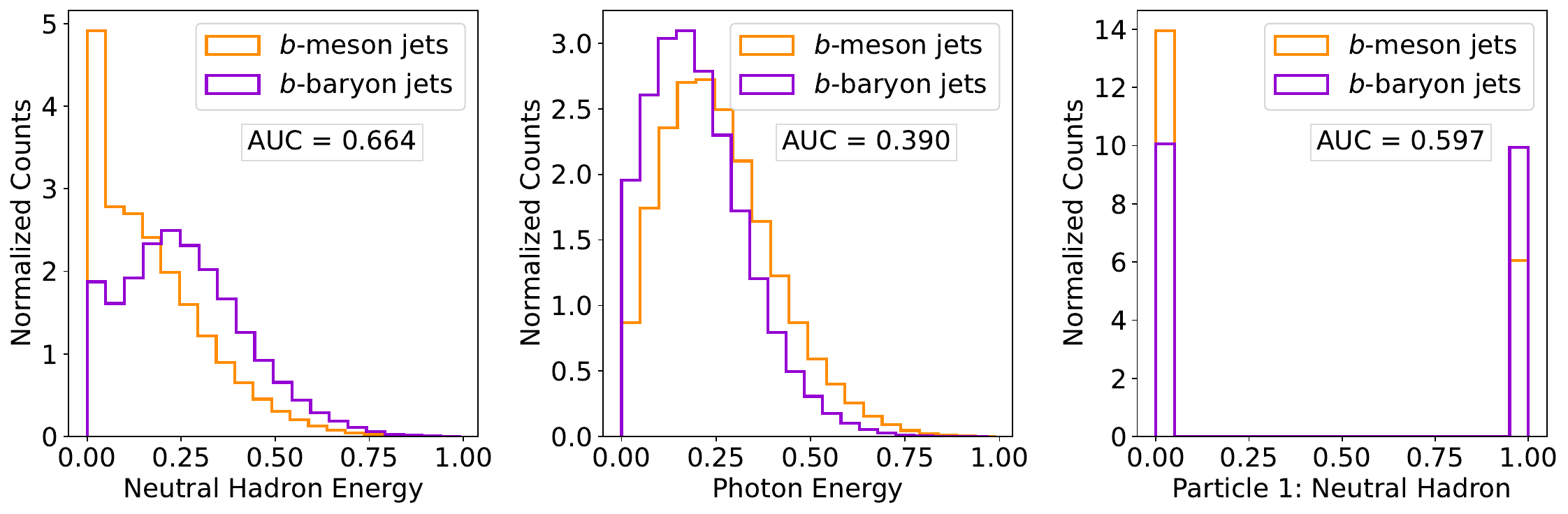}
        \caption{$p_{T,{\rm jet}} > 200$~GeV}
        \label{subfig: b_Distribution_3_Discriminative_Jet_Features_200}
    \end{subfigure}
    \vskip\baselineskip %inserts a vertical space equivalent to the normal line spacing, effectively creating a blank line
    \begin{subfigure}[b]{1.0\textwidth}
        \includegraphics[width=\textwidth]{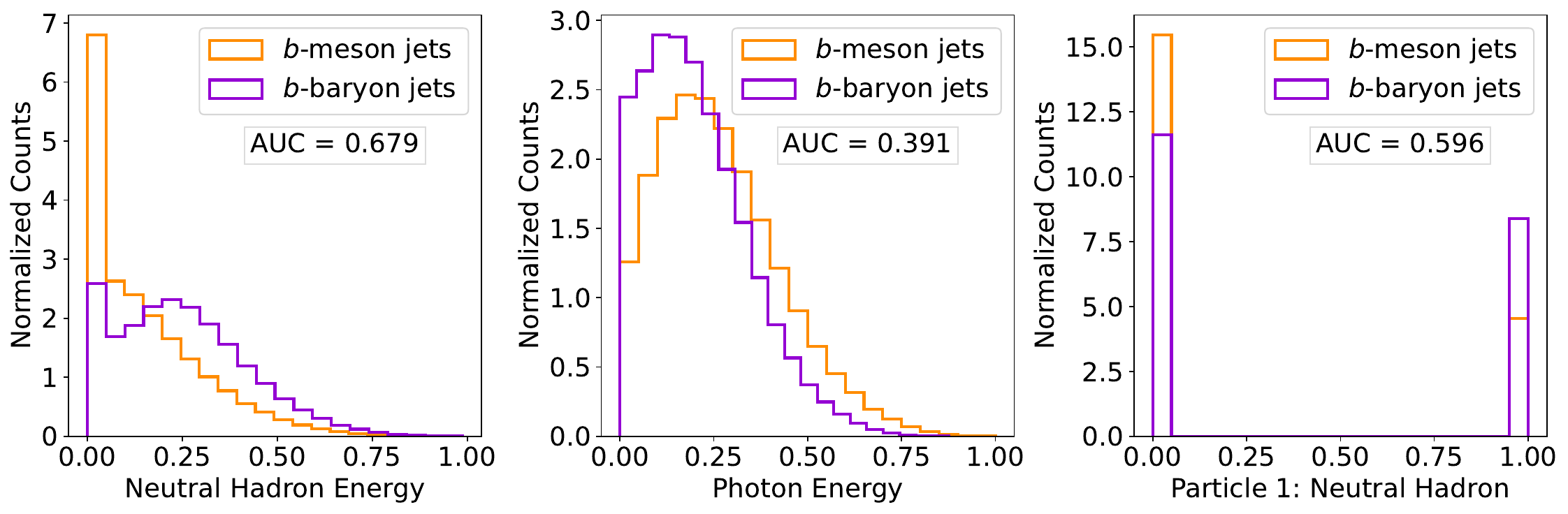}
        \caption{$p_{T,{\rm jet}} > 45$~GeV}
        \label{subfig: b_Distribution_3_Discriminative_Jet_Features_45}
    \end{subfigure}
    \caption{Distributions of the three most discriminative features from figure~\ref{fig: b_ROC For Most Discriminative Jet Features} in $b$-baryon and $b$-meson jets for (a) $p_{T,{\rm jet}} > 200$~GeV, (b) $p_{T,{\rm jet}} > 45$~GeV.}
    \label{fig: b_Distribution_3_Discriminative_Jet_Features}
\end{figure}

Lastly, we construct ROC curves for the different features and compute the resulting AUC scores. Figure~\ref{fig: b_ROC For Most Discriminative Jet Features} presents the ten most discriminative features. The neutral hadron energy and photon energy, along with the identity of the most energetic particle in the jet (Particle 1), exhibit the highest/lowest AUC values. For low-$p_T$ jets (figure~\ref{fig: b_ROC-45 For Most Discriminative Jet Features}), the energy in reconstructed $\Lambda$ baryons is the strongest discriminator for $\varepsilon_s \lesssim 14\%$. The efficiency here is limited by the probability for the jet to contain a $\Lambda$ baryon and for its highly displaced $\Lambda \to p\pi^-$ decay to be reconstructed in the tracker. This discriminator is much less useful for high-$p_T$ jets because the probability for the $\Lambda$ to decay sufficiently early in the tracker becomes too small. Figure~\ref{fig: b_Distribution_3_Discriminative_Jet_Features} shows the distributions of the three most discriminative features identified in figure~\ref{fig: b_ROC For Most Discriminative Jet Features}.

\FloatBarrier

\section{ML-based taggers}
\label{sec: ML}

We will now describe the format of the data that will be fed into the NNs and the architectures we will be using (implemented with PyTorch~\cite{PyTorch}), and then analyze the performance of the taggers.

\subsection{NN inputs}
\label{subsec: NN inputs}

\paragraph{Jet properties}
The whole-jet properties, as described in section~\ref{subsec: jet clustering}, are the jet's $p_T$, $\eta$,\footnote{Since we designed the samples to have similar jet $p_T$ and $\eta$ distributions, these two features are not useful for discrimination. We still provide them to the NNs since their values can be useful for interpreting some of the other features, whose distributions have some dependence on the jet $p_T$ and $\eta$.} number of constituents $N$, and the energy fractions carried by the different types of constituents: photon energy $E_{\gamma}$, electron energy $E_e$, muon energy $E_{\mu}$, charged hadron energy $E_{\text{CH}}$ (in the $b$-jets case, we use two separate variables: charged hadron energy from the primary vertex $E_{\text{CH,PV}}$ and charged hadron energy from the secondary vertex $E_{\text{CH,SV}}$), neutral hadron energy $E_{\text{NH}}$, reconstructed $K_S$ energy $E_{K_S}$, and reconstructed $\Lambda$ energy $E_{\Lambda}$ (including $\overline\Lambda$). The jet's $p_T$, $\eta$, and the number of constituents $N$ are shifted and scaled to have a mean of $0$ and a standard deviation of $1$. This is done to ensure that all features are within a similar range, which is beneficial for the stability and convergence of the machine learning algorithms.

\paragraph{Constituent properties}
The constituent properties, as discussed in section~\ref{subsec: jet clustering}, are the normalized transverse momentum ($p_{T,i}^{\rm norm}$), the angular position relative to the jet axis in terms of $r$ and $\alpha$, and the identity of the constituent. The identity is represented by a set of discrete variables, corresponding to photons, electrons, muons, charged hadrons (in the case of the $b$-jets analysis, there are separate entries for charged hadrons from the primary and those from the secondary vertex), neutral hadrons, and reconstructed $K_S/\Lambda$ particles. For each constituent, the corresponding entry is set to $1$ if it is a positively charged particle, a neutral hadron, or a reconstructed $\Lambda$ baryon, and $-1$ if it is a negatively charged particle or a reconstructed $K_S$ meson, while the other entries are set to $0$.

\paragraph{Graph representation}
We represent each jet as a graph, implemented with Deep Graph Library (DGL)~\cite{DBLP:journals/corr/abs-1909-01315}. The graph nodes represent the jet constituents. Each node's features include the properties of both the jet and the constituent. We employ a fully connected topology, where edges are formed between every pair of nodes. For an edge between node $i$ and node $j$, a vector is initialized with a list of the jet properties, the constituent properties from node $i$, and the constituent properties from node $j$. We leave it to each of the NNs to construct useful edge features based on these physical inputs. Reverse edges (between nodes $j$ and $i$), are included as well, with the order of the nodes in the vector swapped, to allow the independent flow of information in each direction. Each graph in our simulated dataset carries a label to denote the jet type: `0' for $d$-quark jets or $b$-meson jets, and `1' for $s$-quark jets or $b$-baryon jets.

\paragraph{Datasets}
Our strange-tagging datasets contain about one million jets, equally distributed between $s$-quark and $d$-quark jets. Our fragmentation-tagging datasets contain about one million $b$ jets, with a distribution of $30\%$ $b$-baryon and $70\%$ $b$-meson jets (after we discarded a large fraction of the meson jets to avoid a bigger imbalance between the two classes due to the natural rarity of baryons). The datasets are split into training ($72\%$), validation ($18\%$), and testing ($10\%$) samples.

\subsection{NN architectures}

One architecture we use is a variant of a Graph Attention Network (GAT)~\cite{Velickovic:2017lzs,Brody:2021dbs}, a type of Graph Neural Network~\cite{shlomi2020graph,DeZoort:2023vrm}. In this architecture, the node features are updated iteratively with aggregated features of all other nodes, with weights determined through an attention mechanism. In the first iteration, the aggregation weights are determined by an embedding of the physical features of the two nodes and the jet features, as described above. In subsequent iterations, the updated features of the node pairs are used to determine the weights. Finally, the features of all nodes are aggregated and processed to produce the classifier output---a number between 0 and 1. The full structure of this NN is described in appendix~\ref{app: GAT details}.

The most sophisticated NN architecture we consider is based on the idea of the Particle Transformer (ParT), introduced in ref.~\cite{qu2022particle} and inspired by the famous transformer architectures~\cite{vaswani2017attention,DBLP:journals/corr/abs-2103-17239}. Its central element is the \emph{scaled dot-product attention mechanism}, used in two ways. First, in \emph{particle attention blocks}, learned linear projections are applied to each node's features to produce three vectors: \emph{query} ($Q$), \emph{key} ($K$), and \emph{value} ($V$). Each node sends its query to all other nodes. The other nodes respond with their values, with a weight that depends on the similarity (dot product) between the query and their key, as well as on the edge features. (Different from ref.~\cite{qu2022particle}, the edge features in our implementation are not hand-crafted physical quantities but are instead generated by the NN based on the physical properties of the two particles and the jet, as mentioned above.) The original node features are then updated based on these weighted values. This process is repeated several times. Each iteration enhances the information carried by each node as a result of its interactions with the other nodes in the graph. Moreover, several sets of queries, keys, and values, known as \emph{heads}, operate in parallel, implementing \emph{multihead attention}. After the particle attention blocks are completed, a \emph{class token}---an additional node that does not represent any particle---is introduced. In \emph{class attention blocks}, the class token sends queries to all nodes in the graph and, based on the returned weighted values, develops an understanding of the jet as a whole. This procedure is also repeated several times. The class token features are eventually processed to produce the ParT output. For many additional details, see appendix~\ref{app: ParT details}.

We also implement the simplest possible NN architecture---a Multilayer Perceptron (MLP)---that is only given the whole-jet properties and the properties of the most energetic constituent. The purpose of including this architecture alongside more complex models like the GAT and ParT is to serve as a baseline for performance comparison. By evaluating how well these sophisticated NN architectures, which are given the properties of all constituents, perform against the MLP, we can see whether the tasks in question actually benefit from the architectural advantages of these more complicated models. The details of our MLP are given in appendix~\ref{app: MLP details}. 

\subsection{Performance}

We now present the results obtained with each of the tagger types for each of the classification tasks.

\subsubsection{Strange tagging}
\label{subsec: results for s/d jets}

\begin{figure}
    \centering
    \begin{subfigure}[b]{1.0\textwidth}
        \includegraphics[width=\textwidth]{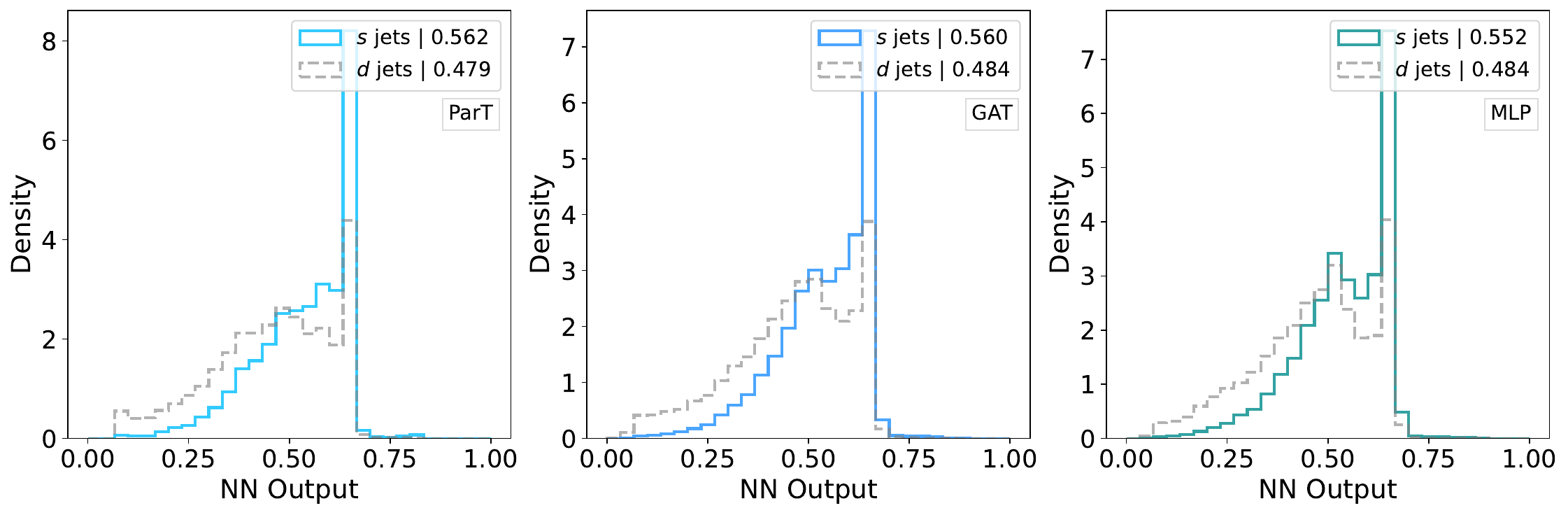}
        \caption{$p_{T,{\rm jet}} > 200$~GeV}
    \end{subfigure}
    \vskip\baselineskip %inserts a vertical space equivalent to the normal line spacing, effectively creating a blank line
    \begin{subfigure}[b]{1.0\textwidth}
        \includegraphics[width=\textwidth]{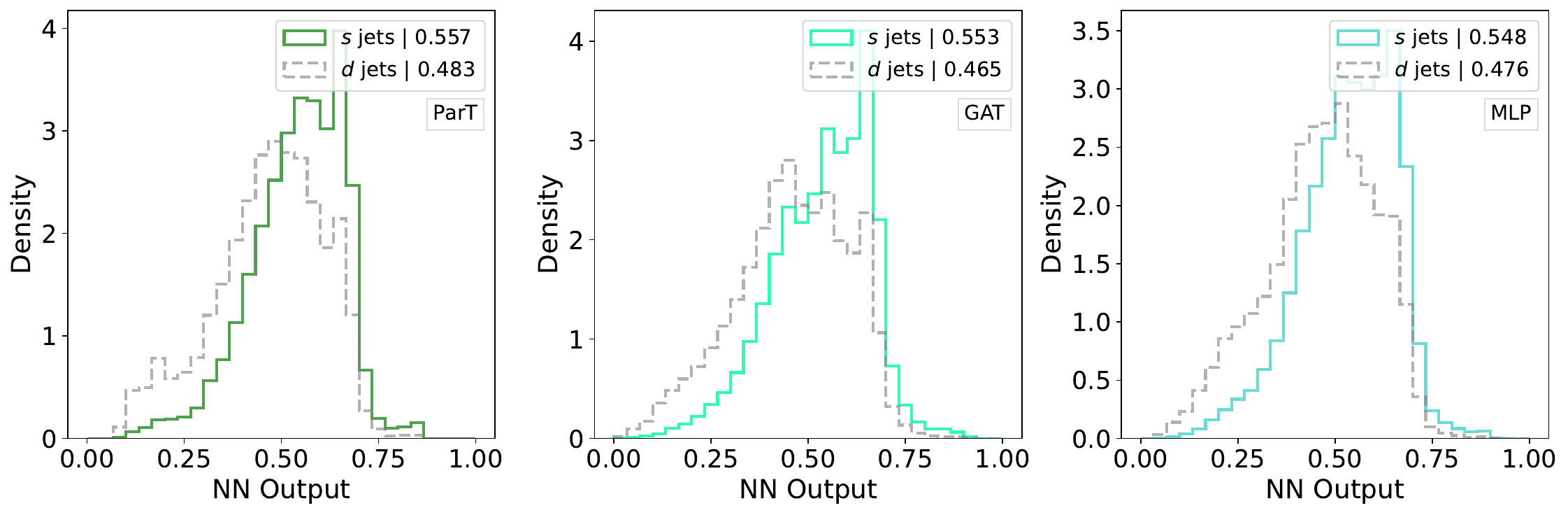}
        \caption{$p_{T,{\rm jet}} > 45$~GeV}
    \end{subfigure}
    \caption{Distributions of the NN outputs for $s$-quark and $d$-quark jets for (a) $p_{T,{\rm jet}} > 200$~GeV, (b) $p_{T,{\rm jet}} > 45$~GeV, for ParT (left), GAT (middle) and MLP (right). The median value for each distribution is indicated in the legend.}
    \label{fig: s_d_Sigmoid_for_the_GNN}
\end{figure}

Figure~\ref{fig: s_d_Sigmoid_for_the_GNN} presents the distributions of the NN outputs for the test datasets. The overlapping distributions show that it is challenging for all the models to clearly distinguish between $s$-quark and $d$-quark jets.

\begin{figure}
    \centering
    \begin{subfigure}{0.45\textwidth}
        \includegraphics[width=\textwidth]{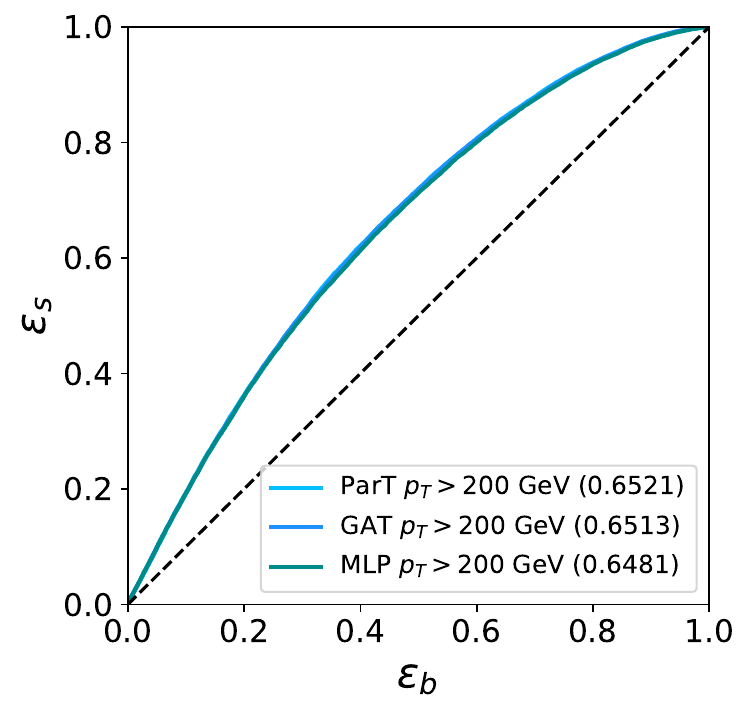}
        \caption{$p_{T,{\rm jet}} > 200$~GeV}
    \end{subfigure}
    % \vskip\baselineskip %inserts a vertical space equivalent to the normal line spacing, effectively creating a blank line
    \begin{subfigure}{0.45\textwidth}
        \includegraphics[width=\textwidth]{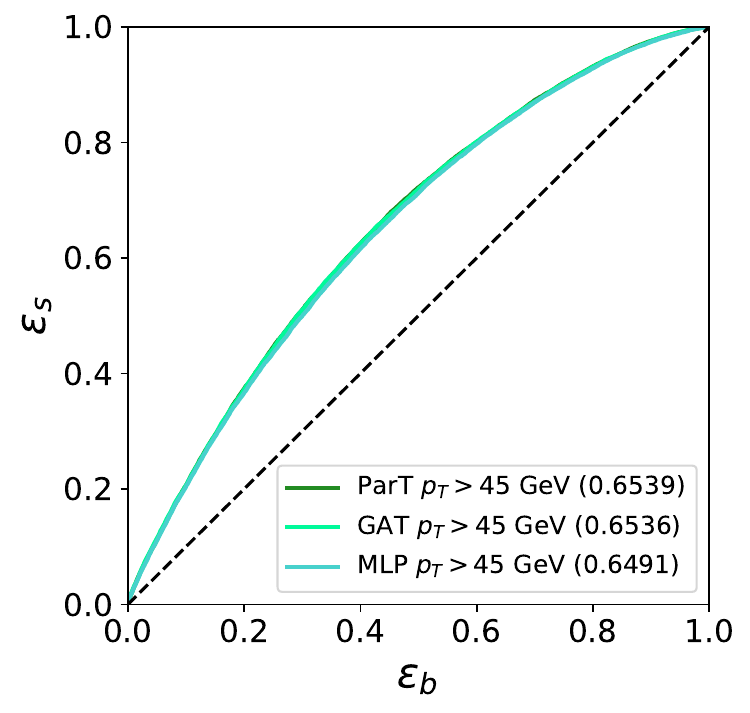}
        \caption{$p_{T,{\rm jet}} > 45$~GeV}
    \end{subfigure}
    \caption{The strange-tagging ROC curves for the different architectures we have used: ParT, GAT, and MLP, for (a) $p_{T,{\rm jet}} > 200$~GeV, and (b) $p_{T,{\rm jet}} > 45$~GeV. The plots show the signal efficiency ($\varepsilon_s$), which is the fraction of $s$-quark jets passing the threshold on the NN output, as a function of the background efficiency ($\varepsilon_b$), indicating the fraction of $d$-quark jets incorrectly identified as $s$-quark jets by the model. The AUC values are given in parentheses in the legends.}
    \label{fig: s_d_ROC_for_the_GNN}
\end{figure}

\begin{figure}
    \centering
    \begin{subfigure}{0.45\textwidth}
        \includegraphics[width=\textwidth]{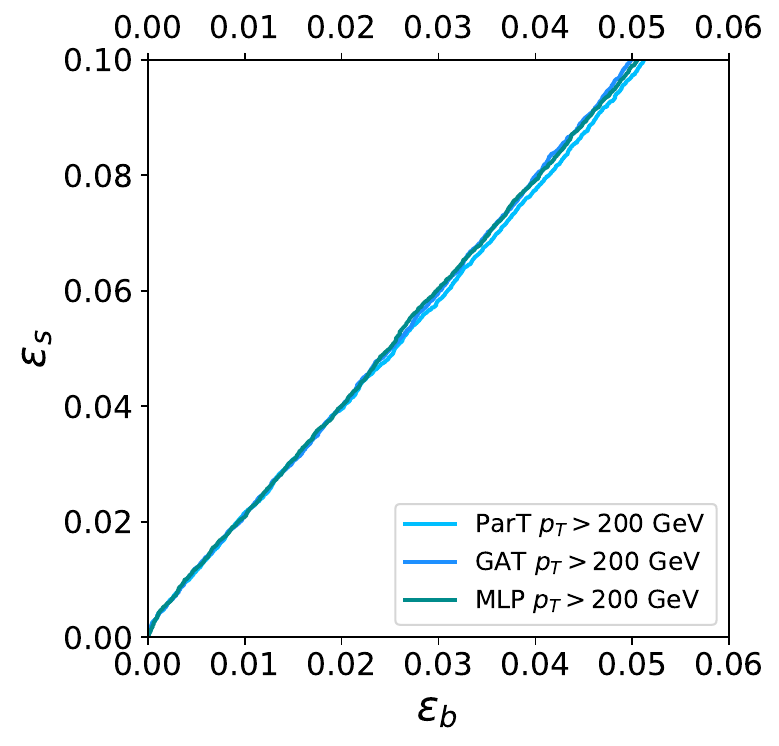}
        \caption{$p_{T,{\rm jet}} > 200$~GeV}
    \end{subfigure}
    % \vskip\baselineskip %inserts a vertical space equivalent to the normal line spacing, effectively creating a blank line
    \begin{subfigure}{0.45\textwidth}
        \includegraphics[width=\textwidth]{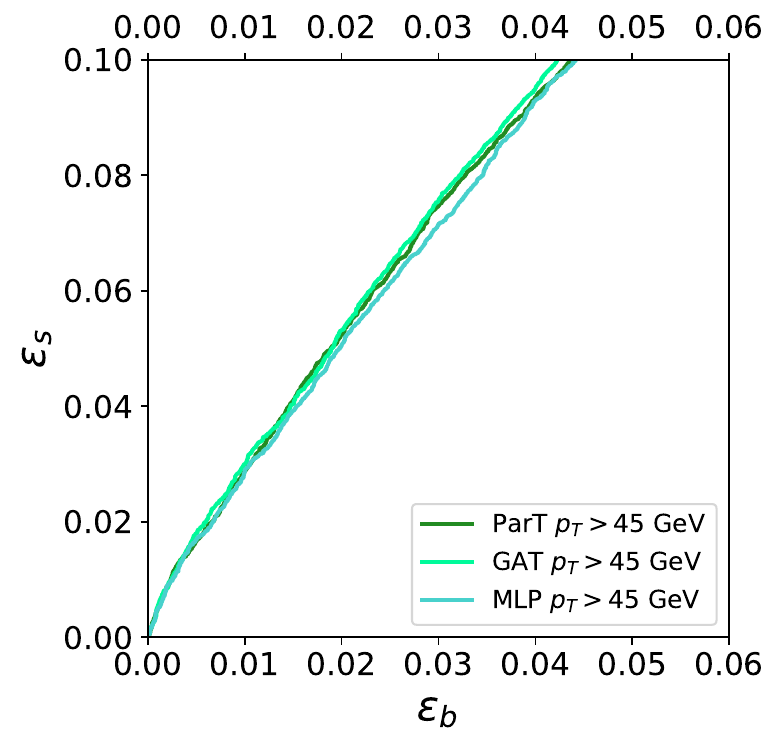}
        \caption{$p_{T,{\rm jet}} > 45$~GeV}
    \end{subfigure}
    \caption{Zoomed-in ROC curves of figure~\ref{fig: s_d_ROC_for_the_GNN} for $s$-quark vs.\ $d$-quark jet classification by the NN models for (a) $p_{T,{\rm jet}} > 200$~GeV, (b) $p_{T,{\rm jet}} > 45$~GeV.}
    \label{fig: s_d_ROC_for_the_GNN_zoomed}
\end{figure}

The ROC curves characterizing the performance of the different models are presented in figure~\ref{fig: s_d_ROC_for_the_GNN}. Figure~\ref{fig: s_d_ROC_for_the_GNN_zoomed} zooms in on the region of low (although still sizable and relevant) signal efficiencies. We can observe that if we choose, for example, to accept $\varepsilon_s = 10\%$ of the signal events, the background efficiency is $\varepsilon_b \approx 4\%$ for $p_{T,{\rm jet}} > 45$~GeV, and $5\%$ for $p_{T,{\rm jet}} > 200$~GeV. In other words, the taggers improve the $s$/$d$ ratio by a factor of $\sim 2$.

Figure~\ref{fig: s_d_ROC_for_the_GNN} shows that all the NNs outperform the most discriminative individual features (cf.\ figure~\ref{fig: ROC For Most Discriminative Jet Features}). However, all the models have very similar performance, which is also quite similar to what has been obtained with the simpler architectures explored in the past: BDTs~\cite{nakai2020strange}, CNNs~\cite{nakai2020strange}, LSTM RNNs~\cite{erdmann2021maximum,zeissner2021development} and FNNs~\cite{zeissner2021development}. The sophisticated GAT and ParT architectures do not bring any improvement in performance, suggesting that the jet data does not contain much useful information beyond what can be captured by a simple combination of hand-crafted variables. This leads us to believe that achieving significantly better strange-tagging performance with the ATLAS and CMS detectors is unlikely, at least with the physical inputs that we assumed to be available and potentially relevant. We note that the significantly better performance expected at the FCC-ee (e.g., ref.~\cite{Blekman:2024wyf}) relies to a large extent on charged hadron identification capabilities, which allow in particular distinguishing $K^\pm$ from $\pi^\pm$.

\FloatBarrier

\begin{figure}[ht]
    \centering
    \begin{subfigure}[b]{1.0\textwidth}
        \includegraphics[width=\textwidth]{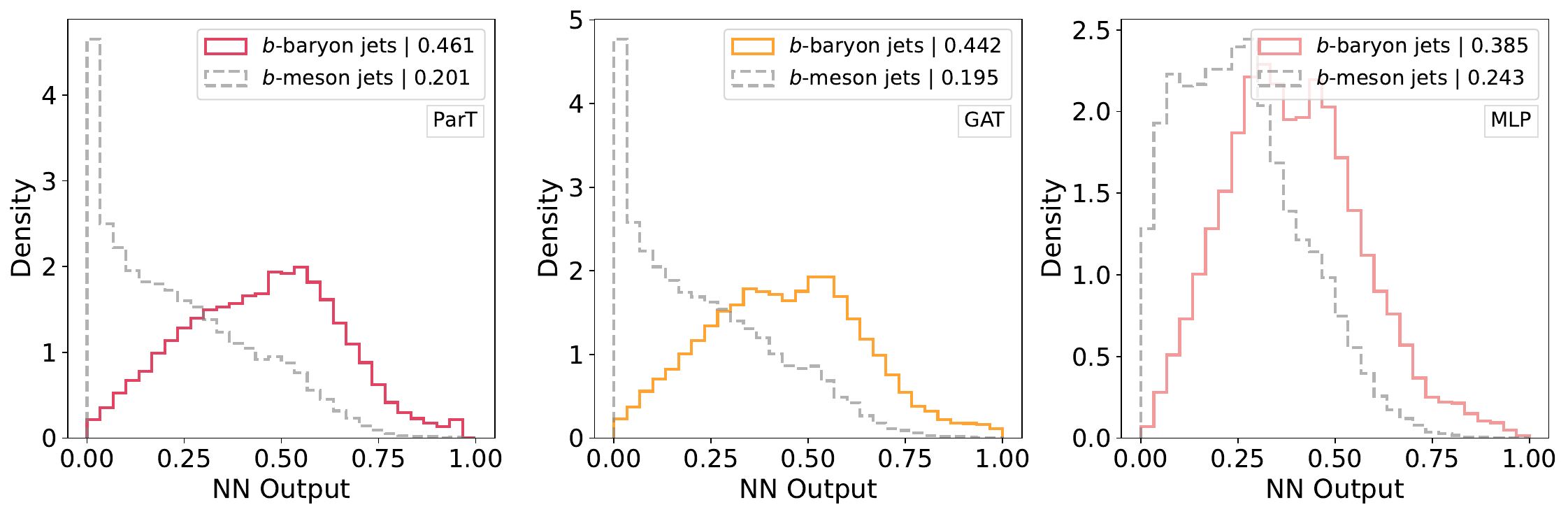}
        \caption{$p_{T,{\rm jet}} > 200$~GeV}
    \end{subfigure}
    \vskip\baselineskip %inserts a vertical space equivalent to the normal line spacing, effectively creating a blank line
    \begin{subfigure}[b]{1.0\textwidth}
        \includegraphics[width=\textwidth]{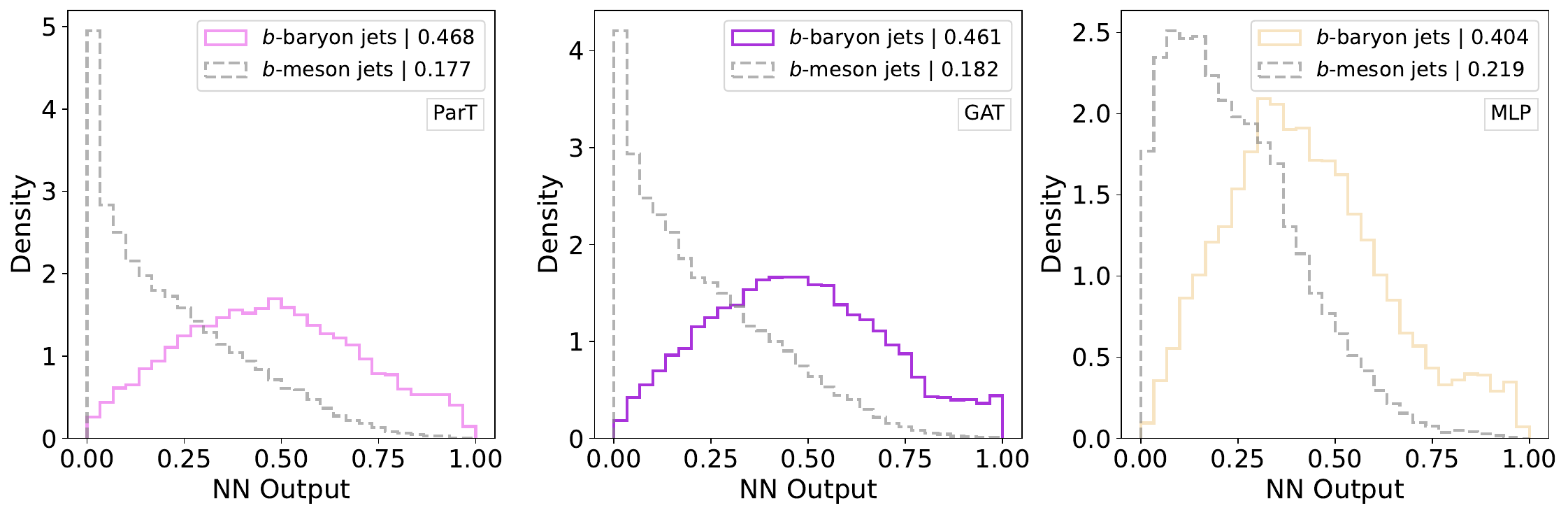}
        \caption{$p_{T,{\rm jet}} > 45$~GeV}
    \end{subfigure}
    \caption{Distributions of the NN outputs for $b$-baryon and $b$-meson jets for (a) $p_{T,{\rm jet}} > 200$~GeV, (b) $p_{T,{\rm jet}} > 45$~GeV, for ParT (left), GAT (middle) and MLP (right). The median value for each distribution is indicated in the legend.}
    \label{fig: bb_Sigmoid_for_the_GNN}
\end{figure}

\subsubsection{Fragmentation tagging}
\label{subsec: results for b jets}

The distributions of the NN outputs from the $b$-baryon/$b$-meson taggers are presented in figure~\ref{fig: bb_Sigmoid_for_the_GNN}. All the models show potential for a decent level of discrimination if one does not insist on having $\mathcal{O}(1)$ efficiencies.

\begin{figure}
    \centering
    \begin{subfigure}{0.49\textwidth}
        \includegraphics[width=\textwidth]{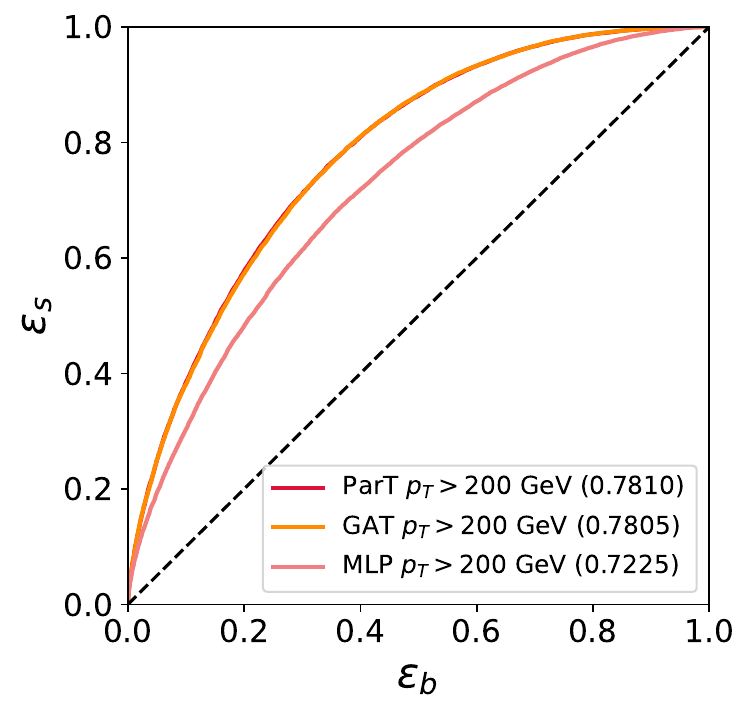}
        \caption{$p_{T,{\rm jet}} > 200$~GeV}
    \end{subfigure}
    % \vskip\baselineskip %inserts a vertical space equivalent to the normal line spacing, effectively creating a blank line
    \begin{subfigure}{0.49\textwidth}
        \includegraphics[width=\textwidth]{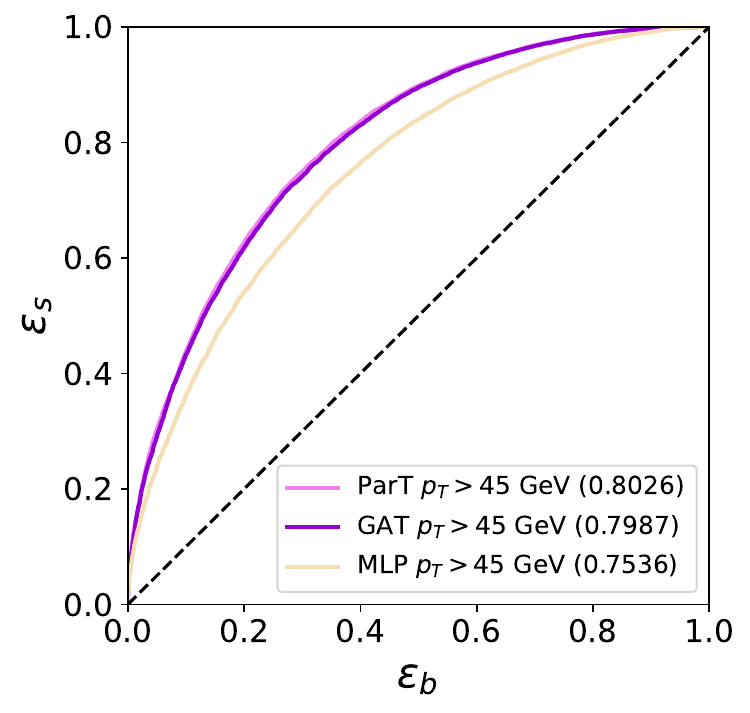}
        \caption{$p_{T,{\rm jet}} > 45$~GeV}
    \end{subfigure}
    \caption{The fragmentation-tagging ROC curves for the different architectures we have used: ParT, GAT, and MLP, for (a) $p_{T,{\rm jet}} > 200$~GeV, and (b) $p_{T,{\rm jet}} > 45$~GeV. The plots show the signal efficiency ($\varepsilon_s$), which is the fraction of $b$-baryon jets passing the threshold on the NN output, as a function of the background efficiency ($\varepsilon_b$), indicating the fraction of $b$-meson jets incorrectly identified as $b$-baryon jets by the model. The AUC values are given in parentheses in the legends.}
    \label{fig: bb_ROC_for_the_GNN}
\end{figure}

The corresponding ROC curves and their AUC values are presented in figure~\ref{fig: bb_ROC_for_the_GNN}, where $b$-baryon jets are taken to be the signal. The GAT and ParT models demonstrate similar performance, which is significantly better than that obtained with the MLP, which in turn is significantly better than that obtained with any individual feature (cf.\ figure~\ref{fig: b_ROC For Most Discriminative Jet Features}).

\begin{figure}
    \centering 
    \begin{subfigure}{0.49\textwidth}
        \includegraphics[width=\textwidth]{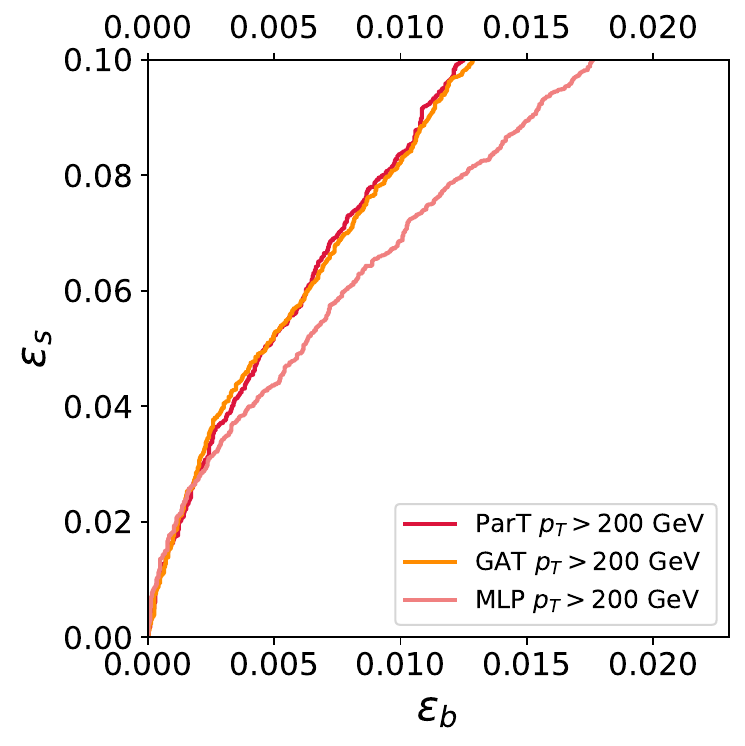}
        \caption{$p_{T,{\rm jet}} > 200$~GeV}
    \end{subfigure}
    % \vskip\baselineskip %inserts a vertical space equivalent to the normal line spacing, effectively creating a blank line
    \begin{subfigure}{0.49\textwidth}
        \includegraphics[width=\textwidth]{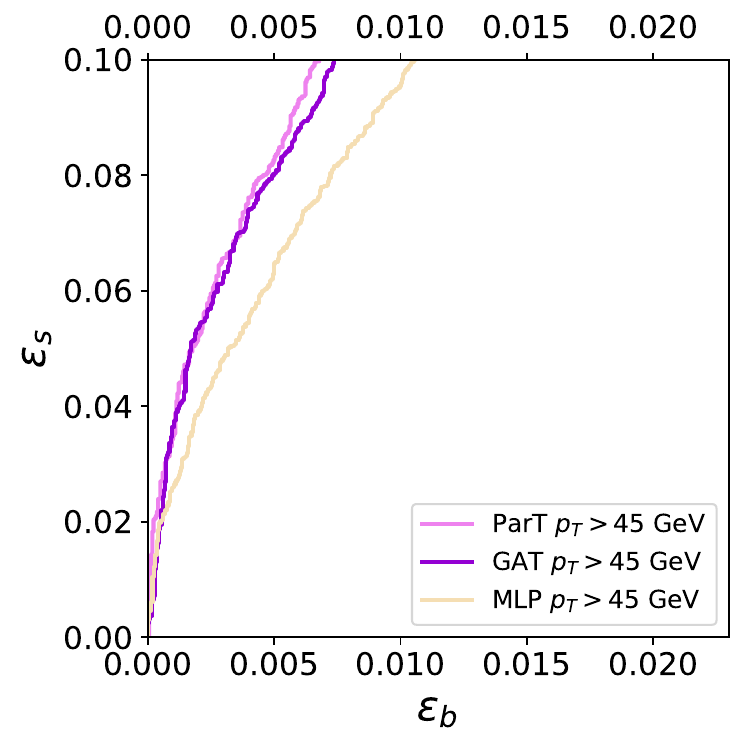}
        \caption{$p_{T,{\rm jet}} > 45$~GeV}
    \end{subfigure}
    \caption{Zoomed-in ROC curves of figure~\ref{fig: bb_ROC_for_the_GNN} for $b$-baryon vs.\ $b$-meson jet classification by the NN models for (a) $p_{T,{\rm jet}} > 200$~GeV, (b) $p_{T,{\rm jet}} > 45$~GeV.}
    \label{fig: b_ROC_for_the_GNN_zoomed}
\end{figure}

We zoom in to low efficiencies in figure~\ref{fig: b_ROC_for_the_GNN_zoomed}. We see that for a signal efficiency of $\varepsilon_s = 10\%$, for $p_{T,{\rm jet}} > 200$~GeV, the ParT and GAT models have a background efficiency of only $\varepsilon_b \approx 1.25\%$, which is better than the MLP by a factor of $1.4$. For $p_{T,{\rm jet}} > 45$~GeV, the ParT model has a background efficiency as low as $\varepsilon_b \approx 0.67\%$, which is slightly better than the GAT, and a factor of $1.6$ better than in the MLP case. Relative to the original samples, the best taggers improve the baryon-to-meson ratio at this signal efficiency by a factor of $8$ in the $p_{T,{\rm jet}} > 200$~GeV case, and $15$ in the $p_{T,{\rm jet}} > 45$~GeV case.

\subsection{Robustness to measurement errors and parton shower modeling}

To assess the robustness of our results against detector resolution effects (which were not initially simulated), we conducted a test by introducing a random $5\%$ measurement error to the transverse momenta of all jet constituents. We then trained our models on this modified data and subsequently tested them on similarly altered data. By comparing the ROC curves and AUC values with and without these induced errors, we observed no significant sensitivity to these effects.

To evaluate sensitivity to parton shower modeling, we performed the following test. We trained the neural networks with the default \textsc{Pythia} parton shower model and then tested their performance on events generated with VINCIA showers~\cite{Fischer:2016vfv}. We found that this did not affect the classification performance noticeably, with changes in the AUC comparable to typical statistical variations of size $0.002$.

\section{Summary and discussion}
\label{sec: summary and discussion}

Jet identification is a broad and imperfectly solved problem in collider physics. In this paper, we addressed two different cases of jet identification at the LHC using variables that are measurable in the ATLAS and CMS detectors.

The first case is \emph{strange tagging}, where we considered the most challenging scenario, which is differentiating between jets originating from strange and down quarks. A strange-quark tagger can in principle be useful for measuring the CKM matrix elements $V_{ts}$ and $V_{td}$ in top-quark decays, and $V_{cs}$ and $V_{cd}$ in $W$ decays. It can also improve the kinematic reconstruction of top-quark decays. Additionally, it can increase the sensitivity to new physics scenarios that involve strange-quark production. Building a strange tagger has been attempted several times in the past, including with machine learning techniques, with a moderate level of success~\cite{erdmann2020tagger,nakai2020strange,erdmann2021maximum,zeissner2021development}. In this work, we approached the same problem with different and more sophisticated neural network architectures.

The second case is \emph{fragmentation tagging}, which in our example differentiates between bottom-baryon jets and bottom-meson jets, but would more generally identify the particular bottom hadron that was produced in the jet. This approach can also be extended to other quark flavors. A fragmentation tagger can be useful for more inclusive measurements of fragmentation functions and for reducing background in the proposed $b$-quark polarization and spin correlation measurements~\cite{galanti2015heavy,kats2023prospects,Afik:2024uif}. To our knowledge, this work is the first attempt at using machine learning for inclusive fragmentation tagging.

The problem of fragmentation tagging is similar to that of discriminating between $s$ and $d$-quark jets. First, in both problems, the pattern of parton showering is the same in the two classes, and therefore cannot be used for discrimination. In addition, the basic properties of the displaced vertex are the same for the different $b$ hadrons, which is analogous to the absence of such a typical displaced vertex in both $s$ and $d$ jets. On the other hand, a remaining handle that can be used in both problems is the different probabilities for the appearance of the various final-state particles in the jet and their kinematics. These stem from the differences in the hadronization processes in the different jets, followed by hadron decays.

In both cases, we structure the jet into a graph format, which represents jet constituents by nodes and uses the properties of each two constituents, combined with the properties of the jet as a whole, as input for edges connecting the corresponding pair of nodes. Such a representation enables the employment of more advanced architectures like the GAT and ParT. These neural networks attempt to identify complex patterns distinguishing between the jets to achieve the desired classification. 

In strange tagging, we found that the GAT and ParT architectures provided with the full jet constituent data did not perform significantly better than a simple MLP that combined the features of the jet and the leading constituent. This extends the result of ref.~\cite{nakai2020strange}, where it was shown that CNNs applied to jet images did not significantly outperform BDTs that used a small number of key whole-jet variables or even a single hand-crafted variable. We summarize our results alongside analogous results from the literature in table~\ref{tab:comparison}.

\begin{table}[t]
\centering
\begin{tabular}{c|c|c|c|c}
Task & Architecture & Study & AUC & $1/\varepsilon_b$ for $\varepsilon_s = 0.1$ \\\hline\hline
\multirow{7}{*}{\makecell{strange \\ tagging}}
  & BDT  & Ref.~\cite{nakai2020strange} & 0.63, 0.63 & 22, 17 \\
  & CNN  & Ref.~\cite{nakai2020strange} & 0.64, 0.64 & 24, 19 \\
  & LSTM & Refs.~\cite{erdmann2021maximum,zeissner2021development} & 0.60 & 14 \\
  & MLP  & Ref.~\cite{zeissner2021development} & 0.61 & 20 \\
  & MLP  & This work & 0.65, 0.65 & 23, 20 \\
  & GAT  & This work & 0.65, 0.65 & 24, 20 \\
  & ParT & This work & 0.65, 0.65 & 23, 20 \\\hline
\multirow{3}{*}{\makecell{fragmentation \\ tagging}}
  & MLP  & This work & 0.75, 0.72 & 95, 57 \\
  & GAT  & This work & 0.80, 0.78 & 140, 80 \\
  & ParT & This work & 0.80, 0.78 & 150, 80 \\
\end{tabular}
\caption{Comparison of performance metrics. For ref.~\cite{nakai2020strange}, the two numbers given in each cell correspond to jets with $p_{T,{\rm jet}} \approx 45$~GeV and jets with $p_{T,{\rm jet}} > 200$~GeV, respectively. Similarly, for our results, the two numbers correspond to jets with $p_{T,{\rm jet}} > 45$~GeV and $p_{T,{\rm jet}} > 200$~GeV, respectively. Small differences in performance between different papers should not be taken too seriously because of differences in sample selection and detector simulation.}
\label{tab:comparison}
\end{table}

Fragmentation tagging, on the other hand, shows promise, especially with the more advanced NN architectures. Our results call for a variety of further studies. These include refining the classification to specific $b$ hadrons, applying the classification to subsets of jets (e.g., those with semileptonic decays), as motivated by particular applications, and extending the framework to other quark flavors. This will likely motivate additional types of input features. Another important question to address is the systematic uncertainties associated with reliance on simulation (in our case, \textsc{Pythia}). It would also be beneficial to develop a scheme to train, or at least calibrate, the classifiers on experimental data. We hope to address some of these questions in future work.

\acknowledgments

This research was supported by the Israel Science Foundation (grant no.~1666/22) and the United States---Israel Binational Science Foundation (grant no.~2018257).

\appendix

\section{NN details}
\label{app: NN details}

This appendix provides the detailed descriptions of the NNs implemented in this work.

\subsection{Graph Attention Network (GAT)}
\label{app: GAT details}

Our Graph Attention Network (GAT) architecture, inspired by refs.~\cite{Velickovic:2017lzs,Brody:2021dbs}, consists of node and edge embeddings, a graph attention block that iterates $n$ times, and a linear block.

\paragraph{Node and edge embeddings}

To transform the physical input features of each node into a more useful representation, we start with Batch Normalization (BN)~\cite{Ioffe:2015ovl}, and then employ three MLP layers with (128, 512, 128) neurons, with each of the layers preceded by Layer Normalization (LN)~\cite{Ba:2016jcy} and followed by the GELU activation function~\cite{Hendrycks:2016qxa}. Similarly, for the edge features, we start with BN and then employ three MLP layers with (32, 64, 16) neurons, applying GELU and LN between the layers, and concluding with BN followed by the sigmoid activation function, which is applied to each of the final layer neurons to produce 16 edge weights within the range $(0,1)$.

\paragraph{Graph attention block}

This block updates each node's features by aggregating the features of all other nodes. The 128 features of each node are divided into 16 groups of 8 features, each corresponding to one of the 16 edge weights. Each weight is applied to aggregate the features within its respective group. The node's own feature vector is also included in the sum, which is then divided by the total number of nodes. The resulting vector is transformed through an MLP block comprising two layers, where the first layer doubles the feature dimension, and the second layer restores it to the original dimension of 128, with GELU and LN applied after the first layer. Subsequently, a residual connection~\cite{he2016deep} followed by a dropout~\cite{Srivastava:2014kpo} is applied.

The edge weights are then recalculated by attending to the updated node features. First, the features of each pair of nodes $i$ and $j$ are concatenated to create new edge features $w_{ij} = [h_i, h_j]$ (where $h_i$ and $h_j$ are the updated node features). These edge features are then transformed through two MLP layers with (64, 16) neurons, with a GELU activation function and LN between the layers, and concluding with the sigmoid activation function applied to each neuron to produce the new weights (attention coefficients).

The entire graph attention block iterates $n = 10$ times.

\paragraph{Linear block}

After the graph attention blocks, the features are averaged across all nodes to form a single vector representing the entire graph, $h_{\text{mean}}$. The features are also summed across all nodes to form another single vector, $h_{\text{sum}}$. We then concatenate these vectors into $h = [h_{\text{mean}}, h_{\text{sum}}]$. This concatenated vector $h$ is processed through a linear layer comprising 64 neurons, followed by a GELU activation function. Subsequently, a second linear layer with a single neuron and the sigmoid activation function produces a number within the range $(0,1)$ as the NN output.

\subsection{Particle Transformer (ParT)}
\label{app: ParT details}

In this section, we review the \emph{scaled dot-product attention mechanism} from ref.~\cite{vaswani2017attention} and the Particle Transformer (ParT) model of ref.~\cite{qu2022particle}.

\paragraph{Attention mechanism}

\begin{figure}
    \centering
    \includegraphics[scale=0.5]{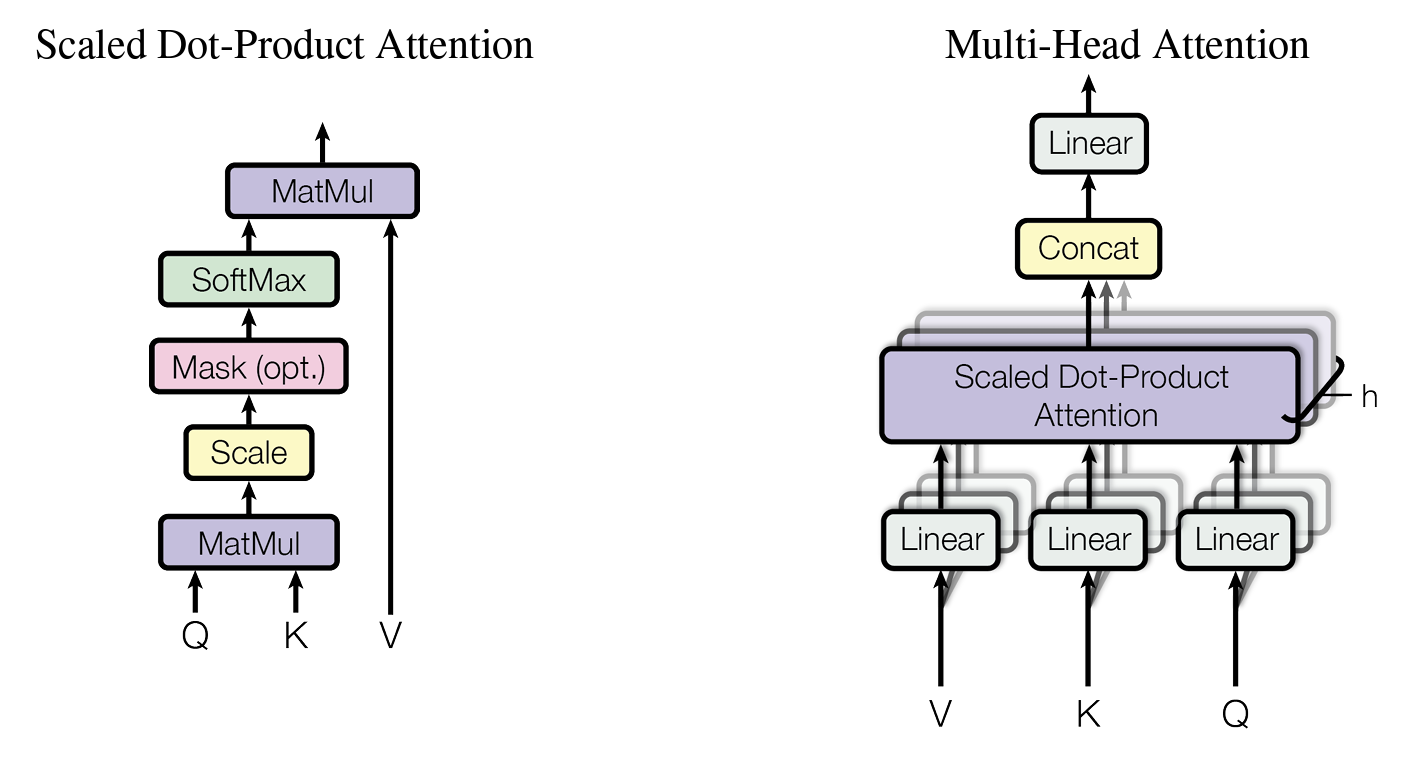}
    \caption{The Scaled Dot-Product Attention (left) and Multi-Head Attention mechanisms (right) from ref.~\cite{vaswani2017attention}.}
    \label{fig: Attention mechanism}
\end{figure}

The scaled dot-product attention mechanism starts by transforming each element (\emph{token}) of the input data through learned linear projections into three vectors: a \emph{query} $Q$, a \emph{key} $K$, and a \emph{value} $V$. Conceptually, the query is like a search term for identifying relevant information, the key acts as a label for each data point, and the value contains the actual data that the mechanism ultimately focuses on based on the computed relevance. Attention scores are computed by calculating the dot product of the query of one token and the key of another, scaling it by $\sqrt{d_k}$,\footnote{The scaling by $\sqrt{d_k}$ is done to prevent the dot product from becoming too large in magnitude (which would force the softmax function to give a number very close to 1), which can cause vanishing gradients during backpropagation~\cite{vaswani2017attention}.} where $d_k$ is the dimensionality of each key and query, and applying the softmax function to obtain the weights that will eventually multiply the values:
\begin{equation}\label{eq: attention_score}
    \text{Attention}(Q,K,V) = \text{softmax}\left(\frac{QK^{T}}{\sqrt{d_k}}\right)V \,,
\end{equation}
as illustrated on the left side of figure~\ref{fig: Attention mechanism}.

The right side of figure~\ref{fig: Attention mechanism} also depicts how the basic attention mechanism is expanded in \emph{Multi-Head Attention (MHA)}. Rather than processing a single set of $Q$, $K$, and $V$, MHA enhances this approach by employing multiple instances of the attention mechanism (``heads'') in parallel enabling the simultaneous processing of the data with several different sets of weights. This design allows the model to focus on various aspects of the input simultaneously. The results from the different heads are then concatenated and linearly transformed to produce the final output.

The variables $Q$, $K$, and $V$ in eq.~\eqref{eq: attention_score} are tensors of dimension $(H, N, d_k)$,\footnote{In general, the feature dimensionality in $V$ can be different from those of $Q$ and $K$, but we will not be using this freedom.} where $H$ is the number of heads and $N$ is the number of tokens (which is the number of nodes in the context of GNNs). The number of features, $d_k$, is taken to be the input dimension divided by the number of heads, to facilitate incorporating residual connections. The dot product (in the feature space for each pair of nodes and each head) $QK^T$ produces a tensor of dimensions $(H, N, N)$. The softmax function is applied over the last dimension of this tensor. The resulting attention scores are multiplied by the value tensor, producing a tensor with dimensions $(H, N, d_k)$. To make this more explicit, we can write eq.~\eqref{eq: attention_score} in components as follows:
\begin{equation}
    \text{Attention}(Q, K, V)_{hni} = \sum_{m} \text{softmax}_{\,m}\left(\frac{1}{\sqrt{d_k}} \sum_{j} Q_{hnj} K_{hmj} \right) V_{hmi} \;.
\end{equation}

\paragraph{Particle Transformer (ParT) architecture}

\begin{figure}
    \centering
    \includegraphics[scale=0.62]{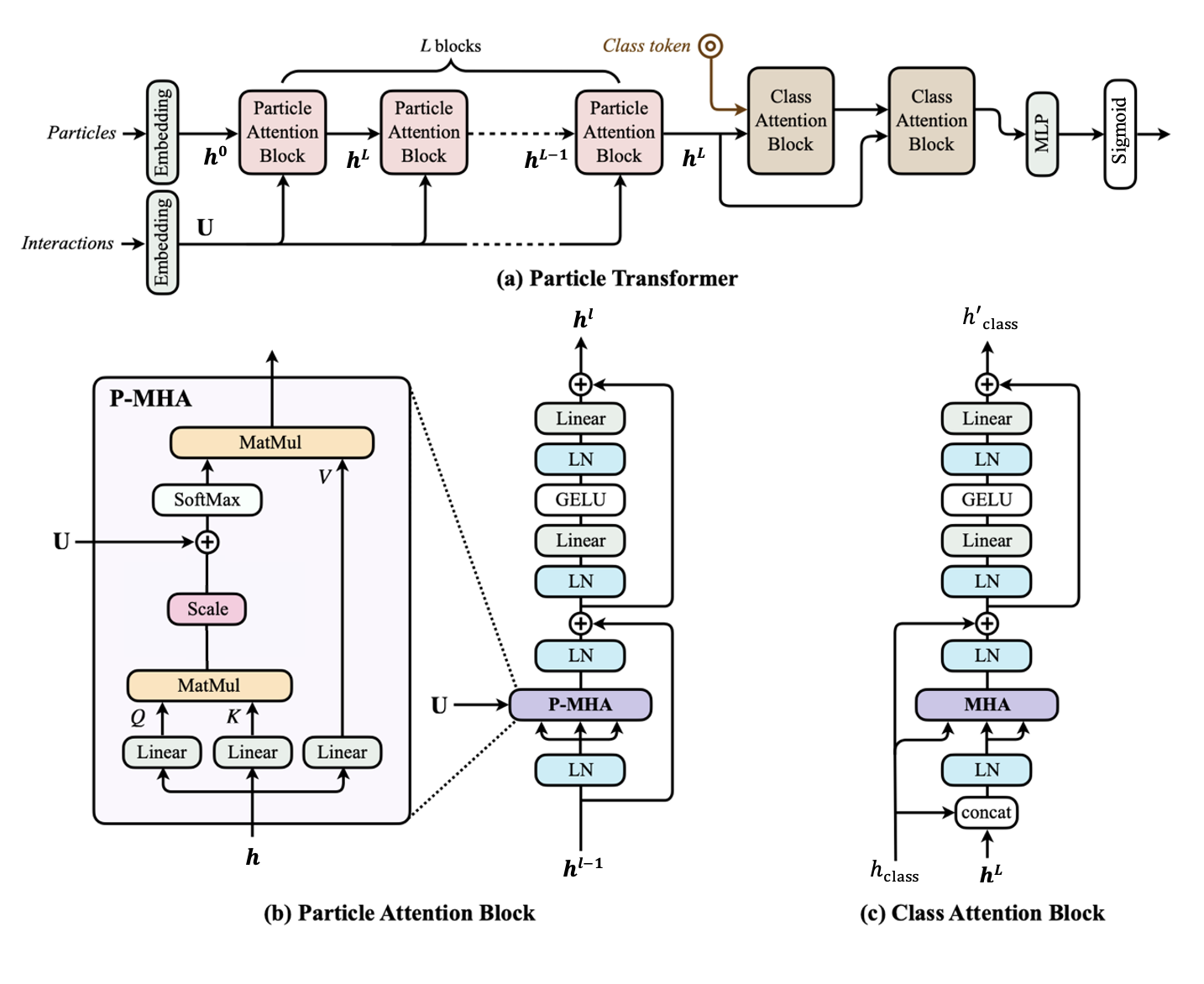}
    \caption{The Particle Transformer (ParT) architecture adapted from ref.~\cite{qu2022particle}. Figure (a) illustrates the overall architecture of ParT, showing how the data is processed through sequential blocks. Figure (b) details the particle attention block, which includes the pairwise multi-head attention (P-MHA) mechanism and linear transformations for feature processing. Figure (c) describes the class attention block, where a class token is integrated to extract global information from the particle nodes via the MHA mechanism and linear transformations.}
    \label{fig: Particle_Transformer_architecture}
\end{figure}

As illustrated in figure~\ref{fig: Particle_Transformer_architecture}, the architecture is segmented into four parts: node and edge embeddings, particle attention blocks, class attention blocks, and MLP. Firstly, the node and edge features are processed in the node/edge embedding stage. The output subsequently passes through the particle attention block $L$ times. Following that, the output progresses through the class attention block $M$ times. Ultimately, it is processed by the MLP, and after the final linear layer, the sigmoid function is activated to produce a value ranging from 0 to 1. 

\paragraph{Node and edge embeddings}

The embedding procedure maps the physical data into a new vector space, using an MLP for the node features and a convolutional approach for the edge features. The node features, after Batch Normalization (BN)~\cite{Ioffe:2015ovl}, are passed through three linear layers with 128, 512, and 128 neurons, with each of the layers preceded by Layer Normalization (LN)~\cite{Ba:2016jcy} and followed by a GELU activation function~\cite{Hendrycks:2016qxa}. The edge features, after BN, are processed through four layers of pointwise 1D convolution\footnote{While the MLP and the pointwise convolution act similarly, we stick to the methodology presented in~\cite{qu2022particle}.} with 64, 64, 64, and 8 channels, with BN and GELU after each layer, except for the last layer, where GELU is not applied. The resulting edge features for each pair of nodes\footnote{Edges from a node to itself are not included.} form the so-called \emph{interaction matrix} $U$.

\paragraph{Particle attention block}

The particle attention block comprises two main parts, as described in figure~\ref{fig: Particle_Transformer_architecture}(b). The first part consists of the \emph{pairwise multi-head attention} (P-MHA) with LN layers before and after the P-MHA. The P-MHA mechanism is similar to the MHA mechanism described above, except that the pairwise interaction matrix $U$ is incorporated as a bias,
\begin{equation}\label{eq: attention_score_with_u}
        \text{Attention}(Q,K,V) = \text{softmax}\left(\frac{QK^{T}}{\sqrt{d_k}} + U \right)V \,,
\end{equation}
to integrate information about the relationships between nodes. The dimension compatibility between $U$ and $QK^T$ is achieved by ensuring that the number of edge features (after the embedding) aligns with the number of heads. Our P-MHA is configured with 8 heads. The same $U$ is used across all particle attention blocks. A dropout~\cite{Srivastava:2014kpo} is applied after the softmax.

The second part of the particle attention block is constructed from an MLP with two linear layers, each preceded by LN, with a GELU activation function and a dropout between the layers. The first linear layer projects the input into a dimensionality of 512, while the second transforms it back to the original input dimension. 

Residual connections~\cite{he2016deep}, preceded by dropouts, are included after each of these two parts.

The particle attention block is repeated $L=8$ times to deepen the network's ability to learn complex patterns.

\paragraph{Class attention block}

The core concept of the class attention blocks is to generate a global graph-level representation. This is achieved by introducing a \emph{class token} $h_{\text{class}}$~\cite{DBLP:journals/corr/abs-2103-17239}, which is a single node, not corresponding to any particular jet constituent, with the same number of features as the constituent nodes $\bm{h}$. The features of $h_{\text{class}}$ are initialized before the first class attention block by learnable parameters. Subsequently, the class token computes graph-level features by sending queries to the nodes $\bm{h}$ (and to itself).

The structure of the class attention block, which is described in figure~\ref{fig: Particle_Transformer_architecture}(c), is similar to that of the particle attention block, with the same hyperparameters, except that dropouts are not included. The main difference is that instead of the P-MHA, the standard MHA described at the beginning of this section is employed, with $Q$, $K$, and $V$ computed as
\begin{align}\label{eq: class_attention_block_Q_K_V_}
    Q &= W_q\, h_{\text{class}} + b_q \,,\\
    K &= W_k\, \bm{z} + b_k \,,\\
    V &= W_v\, \bm{z} + b_v \,.
\end{align}
Here, $\bm{z}=[h_{\text{class}},\bm{h}]$ is the concatenation of $h_{\text{class}}$ and $\bm{h}$, where $\bm{h}$ is the output of the last particle attention block, and $W_i$ and $b_i$ with $i \in \{q, k, v\}$ are learnable weights and biases, respectively.

The class attention block is iterated $M=2$ times. In each iteration, the particle embeddings $\bm{h}$ (the output from the last particle attention block) remain unchanged, while $h_{\text{class}}$ is updated. This process allows the class token to iteratively extract information from the particle embeddings through multiple class attention blocks.

\paragraph{Final MLP layers}

The final stages of the model comprise two linear layers to process the output from the class attention blocks and generate the final prediction. The representation obtained from the class attention blocks is passed through a linear layer, followed by a GELU activation function, to transform it into a lower-dimensional space with 64 neurons. This is followed by a second linear transformation, reducing the dimensions to a single number. Finally, the sigmoid activation function is applied, which brings the output to the range $(0,1)$, forming the model prediction.

\subsection{Multilayer Perceptron (MLP)}
\label{app: MLP details}

The MLP we use comprises five layers with (128, 256, 512, 64, 1) neurons, with a GELU activation function and LN after each layer, except for the final one, where the sigmoid activation function is used.

\subsection{Hyperparameters}
\label{subsec: hyperparameters}

For all models presented, including the ParT, the GAT, and the MLP, we have standardized the initialization of hyperparameters. The batch size is set to 128. The loss function is Binary Cross-Entropy (BCELoss). All the dropout rates are $0.1$. The Adam optimizer~\cite{kingma2017adammethodstochasticoptimization} is employed with a learning rate of $1 \times 10^{-4}$ and a weight decay of $1 \times 10^{-5}$. A learning rate scheduler is utilized to reduce the learning rate by a factor of 0.1 if there is no improvement in the validation loss for a duration of 10 epochs. The training is ended when there is no improvement in the validation loss over 20 epochs. The inference is done with weights from the point with the lowest validation loss. For strange tagging with $p_{T,{\rm jet}} > 200$~GeV, this point for the ParT, GAT, and MLP models was at 27, 28, and 36 epochs, respectively, while for $p_{T,{\rm jet}} > 45$~GeV, it was at 13, 36, and 29 epochs, respectively. For fragmentation tagging with $p_{T,{\rm jet}} > 200$~GeV, the lowest validation loss for the ParT, GAT, and MLP models was attained at 26, 40, and 41 epochs, respectively, while for $p_{T,{\rm jet}} > 45$~GeV, it was at 31, 29, and 31 epochs, respectively.

\bibliographystyle{utphys}
\bibliography{thesis}

\end{document}